\documentclass[conference]{IEEEtran}
\IEEEoverridecommandlockouts

\usepackage{graphicx}
\usepackage{cite}
\usepackage{booktabs}
\usepackage{amsmath, amssymb, amsfonts}
\usepackage{makecell}
\usepackage{multirow}
\usepackage{algorithm}
\usepackage{algpseudocode}
\usepackage{siunitx}
\usepackage{url}

\graphicspath{{fig/}}

\newcommand{\order}[1]{\mathcal{O}(#1)}

\begin{document}

\title{Design of a Quantum Error Correction Decoder Exploiting Temporal Parallelism}

\author{
  \IEEEauthorblockN{Leo Itoh, Takuya Kasamura, Hidetsugu Irie, and Junichiro Kadomoto}
  \IEEEauthorblockA{The University of Tokyo\\
    \{itoh, kasamura, irie, kadomoto\}@mtl.t.u-tokyo.ac.jp}
  \thanks{Accepted for publication in the 2026 IEEE International Conference
    on Quantum Computing and Engineering (QCE). This is the author's accepted
    version; the final published version will be available via IEEE Xplore.}
  \thanks{\copyright{} 2026 IEEE. Personal use of this material is permitted.
    Permission from IEEE must be obtained for all other uses, in any current or
    future media, including reprinting/republishing this material for advertising
    or promotional purposes, creating new collective works, for resale or
    redistribution to servers or lists, or reuse of any copyrighted component of
    this work in other works.}
}

\maketitle

\begin{abstract}

In the pursuit of fault-tolerant quantum computing, low-latency quantum error correction (QEC) is essential to prevent rapid error accumulation within the syndrome measurement cycle.
In this work, we propose a microarchitecture that implements the sandwich decoding method using the Union-Find algorithm by exploiting temporal parallelism, and we present an ASIC implementation.
Through logic synthesis and simulation, we show that the proposed decoder achieves an average latency reduction of 35\% at code distance d = 21 compared with a conventional batch Union-Find decoder, while maintaining a comparable logical error threshold of approximately 1.5\% under phenomenological noise.

\end{abstract}


\section{Introduction}
Quantum computers, leveraging quantum phenomena such as superposition and entanglement, offer the potential to solve certain classes of computationally intensive problems that are infeasible for classical computers~\cite{daley2022jul}.
However, the qubits that underpin quantum computers are more sensitive to noise than classical bits, and errors occur frequently due to disturbances such as thermal noise and measurement.

To realize a quantum computer that is resilient to such frequent errors, i.e., a fault-tolerant quantum computer (FTQC), efficient quantum error correction (QEC) is essential.
In QEC, logical qubits are constructed by encoding physical qubits using appropriate error-correcting codes, such as surface codes~\cite{kitaev2003jan, dennisTopologicalQuantumMemory2002}.
Errors occurring on physical qubits are detected indirectly through syndrome measurements on measurement qubits, and the data qubits on which errors occurred are inferred based on the obtained syndrome information.
This process of estimating errors from syndrome information is called \emph{decoding}.

QEC is performed continuously during quantum computation. If the time required for decoding exceeds the time for syndrome measurement, error accumulation causes the speed of quantum computation to degrade exponentially~\cite{terhal2015apr}.
Simulations of quantum control processors that take these constraints into account indicate that decoding within \SI{1}{\micro\second} per syndrome measurement round is necessary~\cite{byunXQsimModelingCrosstechnology2022}.
This constraint originates from the microsecond-scale syndrome measurement cycle in superconducting quantum processors.
To meet this requirement, a high-speed QEC decoder implemented in dedicated hardware is needed.

To improve decoding speed, methods that exploit temporal parallelism (performing decoding incrementally in parallel with syndrome measurements rather than waiting for the complete measurement history) have been proposed~\cite{tanScalableSurfaceCode2022, skoricParallelWindowDecoding2023}.
However, these methods remain primarily theoretical proposals, and concrete microarchitectural studies addressing how to implement pipelining and parallelization on actual hardware remain limited.
Furthermore, the extent to which temporally parallel decoding methods reduce latency compared with conventional approaches has not been quantitatively evaluated.

In this work, we propose a parallel-processing microarchitecture capable of executing sandwich decoding, a temporally parallel method.
We adopt the Union-Find algorithm as the internal decoder, implement it targeting ASIC fabrication, and evaluate its performance.
As a result, we show that the proposed method reduces average latency by \SI{35}{\percent} and worst-case latency by \SI{12}{\percent} compared with a conventional batch decoder at code distance $d=21$.
We also confirm that the threshold of the sandwich decoder is approximately \SI{1.5}{\percent} under phenomenological noise, which, although slightly lower than that of the batch decoder, maintains comparable performance.

\section{Background}

\subsection{Surface Codes}

\begin{figure}[tb]
    \centering
    \includegraphics[width=0.7\columnwidth]{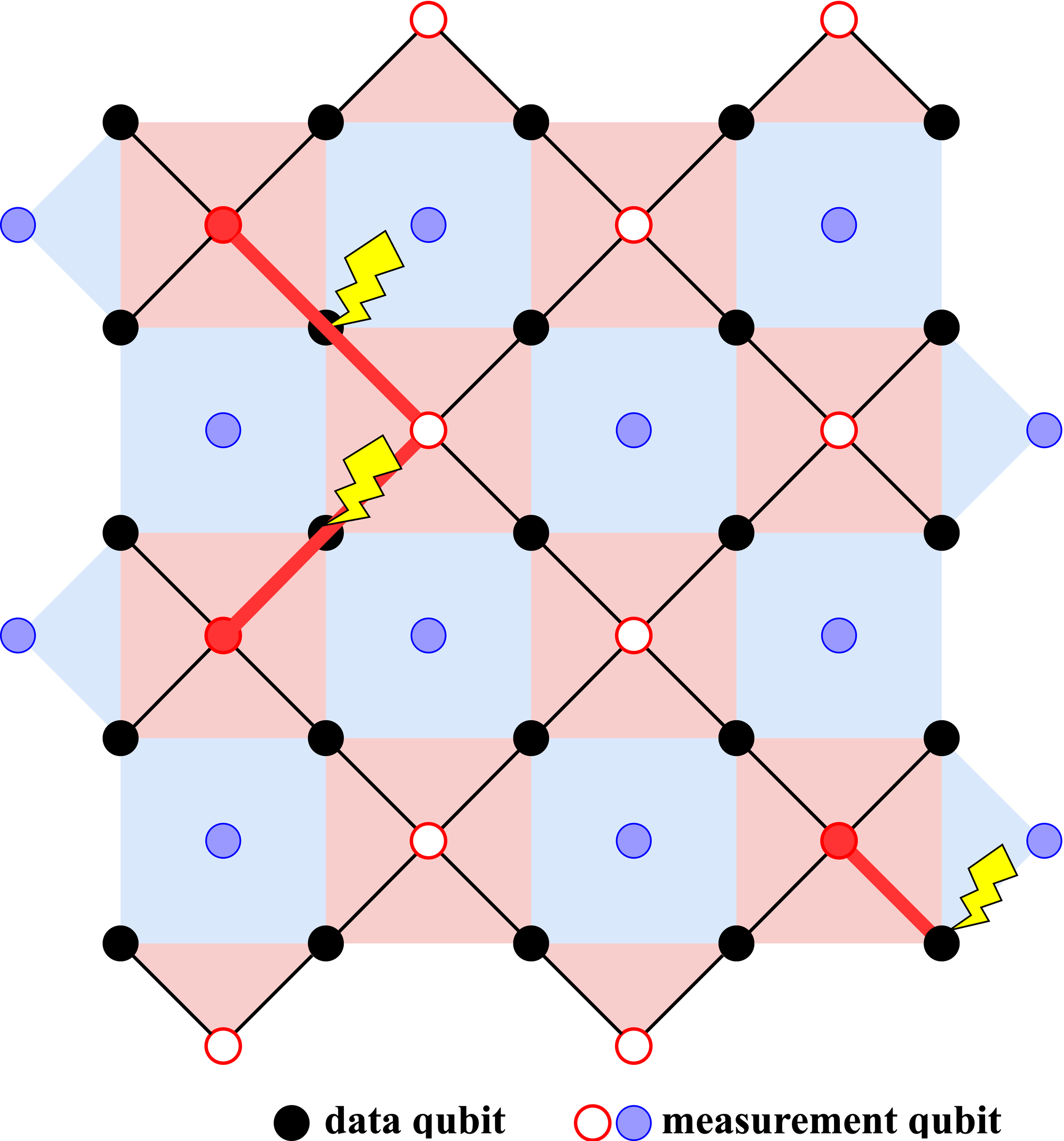}
    \caption{Rotated surface code (distance $d=5$). Data qubits (black) store logical information. Syndrome qubits (red=Z-checks, blue=X-checks) detect adjacent data qubit errors. When an error occurs (lightning), it flips the syndrome of neighboring measurement qubits, enabling error location inference through decoding.}
    \label{fig:rotated_surface_code}
\end{figure}

The surface code is widely used as an efficient quantum error-correcting code for QEC~\cite{kitaev2003jan, dennisTopologicalQuantumMemory2002}.
This code encodes a logical qubit by arranging many physical qubits on a lattice.
The physical qubits are divided into two types: data qubits and measurement qubits.
Data qubits hold the logical state of the code, while measurement qubits are used to indirectly measure the state of adjacent data qubits.

A surface code of distance~5 uses the lattice structure shown in Fig.~\ref{fig:rotated_surface_code}.
Each black dot represents a data qubit, each red dot represents a $Z$-measurement qubit, and each blue dot represents an $X$-measurement qubit.
Each measurement qubit measures the state of four or two adjacent data qubits.
The two types of errors, $X$ and $Z$, that occur on physical qubits change the syndromes of adjacent $Z$-measurement qubits and $X$-measurement qubits, respectively.
Because the syndrome information for these two error types is independent, we consider the case of $X$ errors in the following discussion.

In the surface code, when the physical error rate~$p$ of physical qubits falls below a certain threshold~$p_{th}$, the logical error rate~$p_L$ decreases exponentially with increasing code distance~$d$. 
This property is known as the threshold theorem~\cite{fowler2012sep}.
The logical error rate~$p_L$ is expressed as:
\begin{equation}
    p_L \propto \left( \frac{p}{p_{th}} \right)^{(d+1)/2}
\end{equation}
This threshold depends on the code and the decoding algorithm. Realizing FTQC requires a threshold higher than the physical error rate of the physical qubits.

\subsection{Decoding Surface Codes}

\begin{figure}[tb]
    \centering
    \includegraphics[width=0.7\columnwidth]{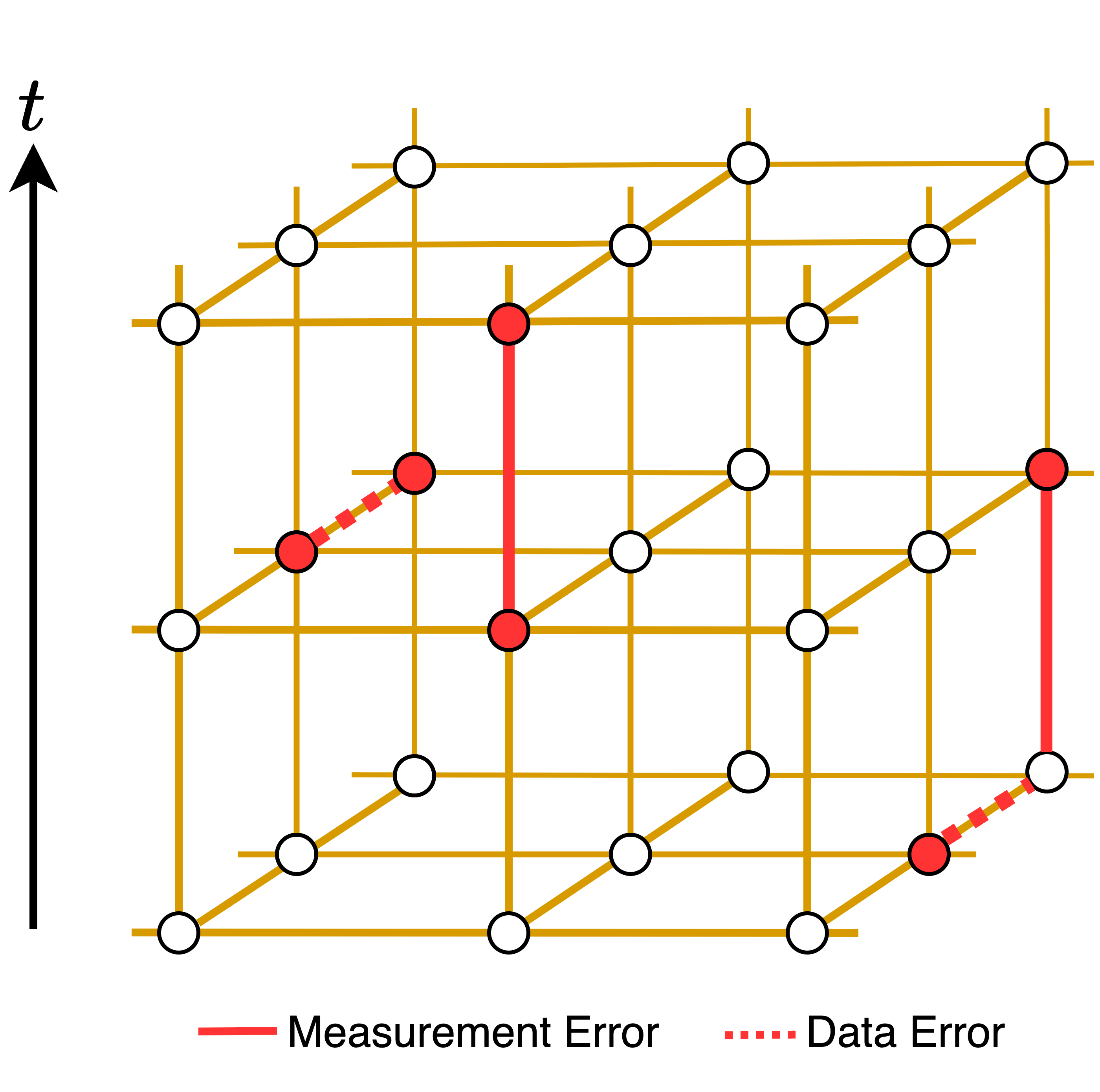}
    \caption{Three-dimensional spatiotemporal decoding graph. Edges along the time axis (solid lines) correspond to measurement errors, while edges along spatial directions (dashed lines) correspond to data qubit errors.}
    \label{fig:decoding_graph}
\end{figure}

In QEC, the errors that must be corrected include not only data qubit errors but also measurement errors on measurement qubits.
Dennis et al.\ proposed a method that reduces such noisy decoding to a graph-matching problem on a three-dimensional spatiotemporal graph and demonstrated its effectiveness~\cite{dennisTopologicalQuantumMemory2002}.

Consider collecting the syndromes of measurement qubits over multiple rounds in a surface code.
Define the detection event $d^{(t)}$ as the exclusive-OR of the syndrome measurement $s^{(t)}$ at round~$t$ and the syndrome measurement $s^{(t-1)}$ at the preceding round~$t-1$:
\begin{equation}
    d^{(t)} = s^{(t)} \oplus s^{(t-1)}
\end{equation}
If no error occurs at round~$t$, the syndrome does not change and $d^{(t)} = 0$.
Conversely, if a data qubit error occurs or a measurement error flips the measurement outcome, a detection event $d^{(t)} = 1$ is triggered on the corresponding measurement qubit.
When these detection events $d^{(t)}$ are placed as nodes on a three-dimensional spatiotemporal lattice (a two-dimensional spatial lattice extended with a time axis), errors are represented as edges connecting pairs of nodes, as shown in Fig.~\ref{fig:decoding_graph}.
Spatial edges correspond to data qubit errors and connect detection events between adjacent measurement qubits within the same round.
Temporal edges correspond to measurement errors and connect detection events at different rounds for the same stabilizer position.
On such a graph, the decoding problem reduces to a matching problem: finding the most likely set of edges (error chains) connecting pairs of detection event nodes on the three-dimensional spatiotemporal graph.

Estimating the most likely error chain for the detected syndrome information constitutes maximum-likelihood decoding.
Because the probability of an error chain~$E$ is the product of the probabilities~$p_e$ of its constituent edges, taking logarithms yields a sum of edge weights:
\begin{equation}
    P(E) = \prod_{e \in E} p_e \quad \Rightarrow \quad \log P(E) = \sum_{e \in E} \log p_e
\end{equation}
Therefore, maximum-likelihood estimation can be formulated as minimizing the sum of edge weights, reducing to the minimum-weight perfect matching (MWPM) problem on a three-dimensional graph~\cite{dennisTopologicalQuantumMemory2002}.
The Blossom algorithm has been proposed to solve such MWPM problems~\cite{kolmogorov2009jul}.
However, this algorithm requires $\order{n^3}$ time complexity for $n$ measurement qubits, posing a challenge for high-speed decoding in large-scale FTQC.
To address this, approximate but fast algorithms such as Greedy~\cite{ueno2021dec}, Tensor Network~\cite{varsamopoulos2020feb}, and Union-Find~\cite{delfosse2021dec} have been proposed~\cite{holmes2020apr}.

\section{Related Work}

\subsection{Union-Find Decoder}

Union-Find is an algorithm for efficiently performing Union (merging sets) and Find (finding which set an element belongs to) operations on a disjoint-set data structure~\cite{galler1964may, tarjan1975apr}.
Delfosse et al.\ proposed the Union-Find decoder, an approximate algorithm for MWPM based on the Union-Find algorithm~\cite{delfosse2021dec, wu2022nov}.

\begin{figure}[tb]
    \centering
    \newlength{\figheightB}
    \setlength{\figheightB}{2.3cm}

    \begin{minipage}[b]{0.33\columnwidth}
        \centering
        \includegraphics[height=\figheightB, keepaspectratio]{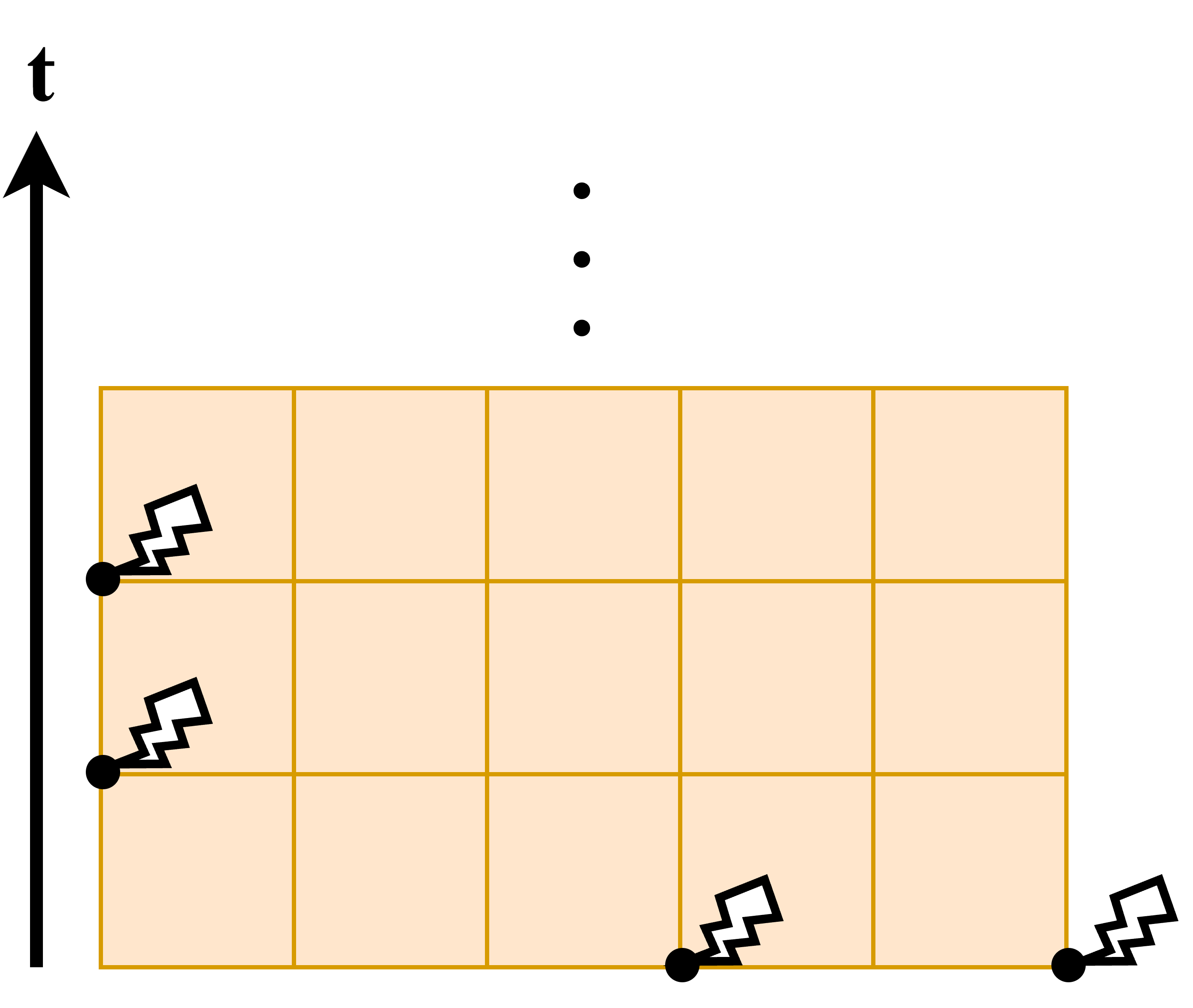}\\
        (a) Measurement.
    \end{minipage}\hfill
    \begin{minipage}[b]{0.33\columnwidth}
        \centering
        \includegraphics[height=\figheightB, keepaspectratio]{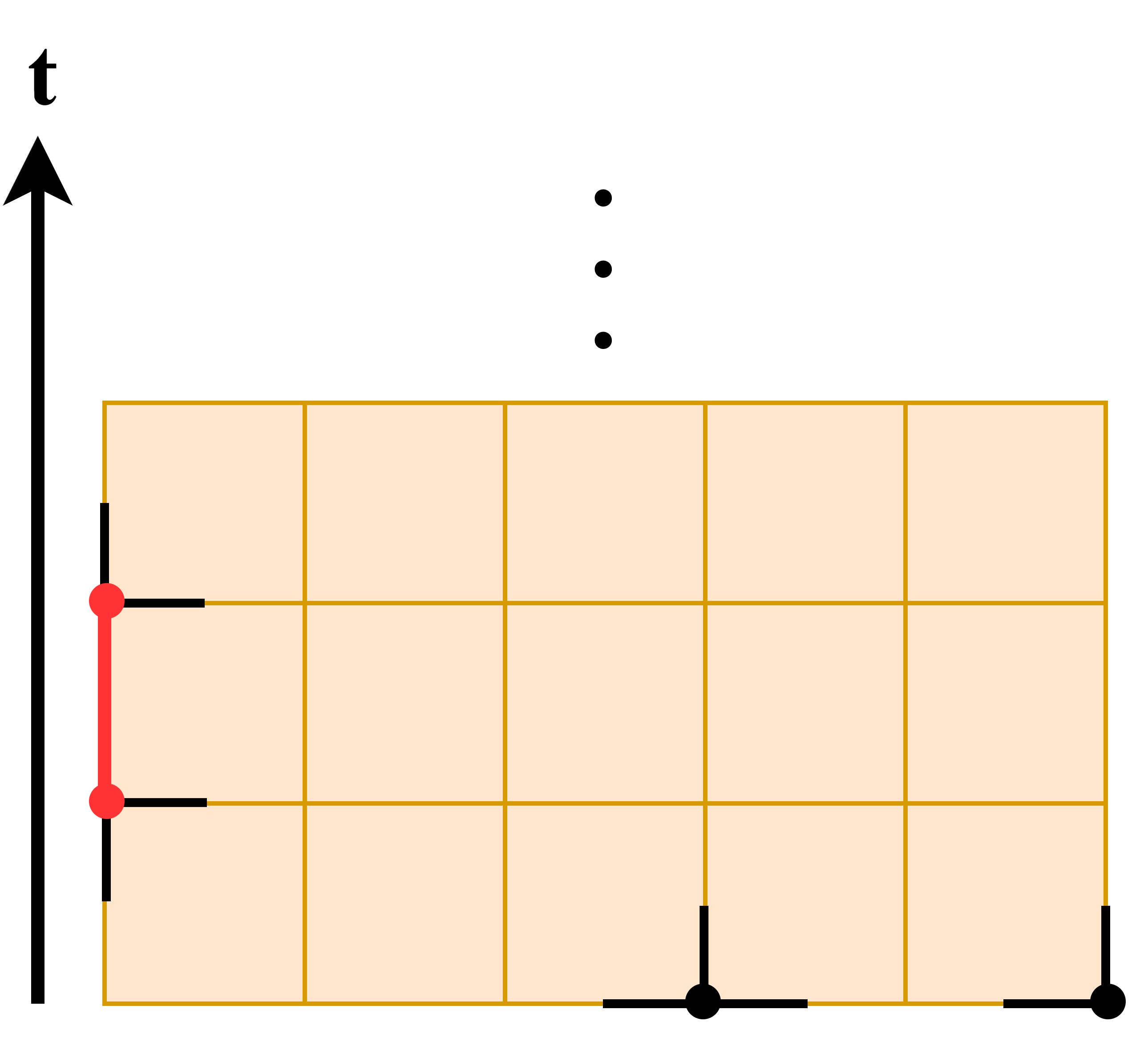}\\
        (b) Grow step 1.
    \end{minipage}\hfill
    \begin{minipage}[b]{0.33\columnwidth}
        \centering
        \includegraphics[height=\figheightB, keepaspectratio]{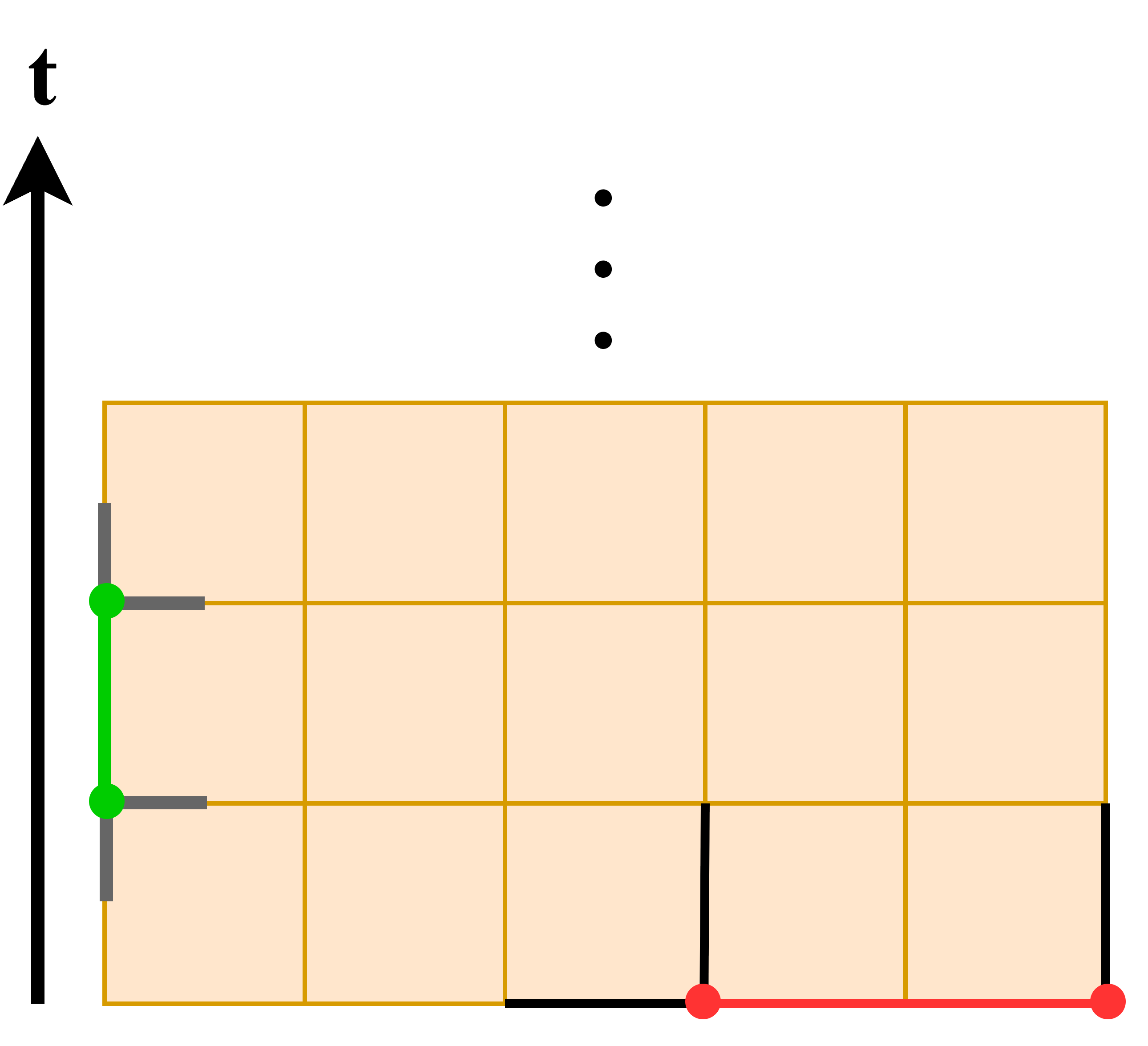}\\
        (c) Grow step 2.
    \end{minipage}

    \caption{Example of cluster growth. The observed syndromes grow by one half-edge at a time, and clusters that come into contact are merged.}
    \label{fig:uf_process}
\end{figure}

In the Union-Find decoder, one error cluster is first generated for each piece of syndrome information, and each cluster is represented as a rooted tree.
Then, as shown in Fig.~\ref{fig:uf_process}, each cluster grows by one half-edge at a time.
When growing clusters come into contact, the Union operation merges the two clusters into one, and the Find operation updates the root.
This process is repeated until all clusters reach even parity.
From the resulting spanning trees, error chains are recovered by tracing edges from the leaf nodes to the root of each tree (\emph{peel}).

Cluster merging and root lookup are performed by the standard \textsc{Union} and \textsc{Find} operations on the disjoint-set forest~\cite{galler1964may, tarjan1975apr}.
In a na\"ive implementation of the Union-Find decoder, the Find operation requires $\order{n}$ time complexity for~$n$ measurement qubits, forming the bottleneck of the entire decoder.
However, by applying optimizations such as path compression and union by size, the amortized time complexity of the Find operation can be reduced to $\order{\alpha(n)}$~\cite{tarjan1975apr}, where $\alpha(n)$ is the inverse Ackermann function, an extremely slowly growing function.
Consequently, the overall time complexity of the Union-Find decoder becomes $\order{n\alpha(n)}$, enabling a significant speedup compared with the MWPM algorithm.

\begin{figure*}[tb]
\centering
\includegraphics[width=0.8\textwidth]{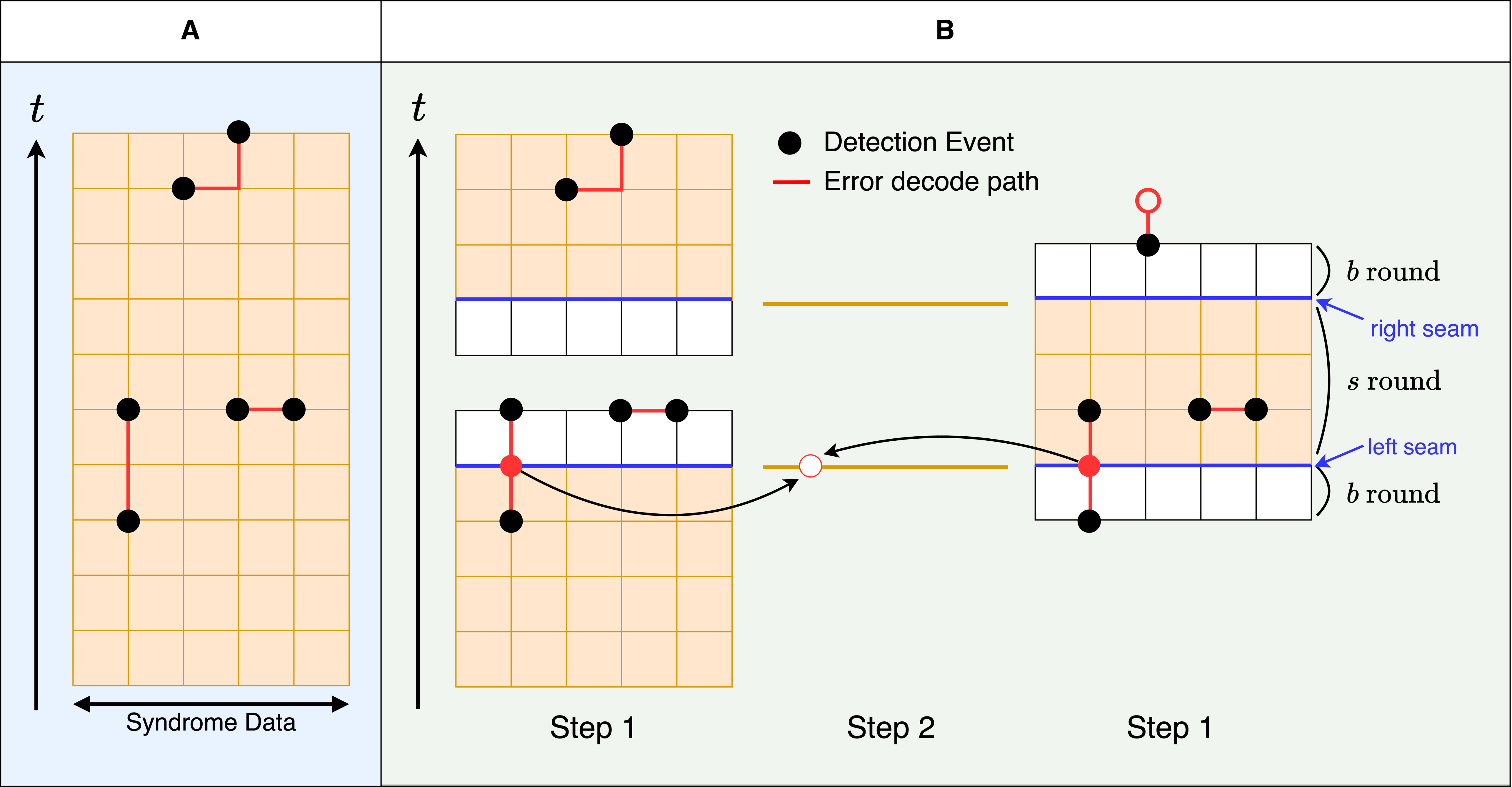}
\caption{Comparison of (A) batch decoding and (B) sandwich decoding on the same syndrome data. Black dots denote detection events, red lines denote estimated error chains, and open circles denote virtual boundary nodes. In batch decoding, $d$ rounds of syndrome measurements are decoded all at once. In sandwich decoding, windows of $s + 2b$ rounds are decoded in parallel (Step~1). An error chain may terminate on a window's time boundary, and such chains are reconciled across the seams (blue lines) between the core regions (shaded) of adjacent windows by the Type-2 decoder (Step~2).}
\label{fig:sandwich_batch}
\end{figure*}

\subsection{Sandwich Decoding}

Most hardware decoders proposed to date adopt a common processing method called batch decoding.
As shown in Fig.~\ref{fig:sandwich_batch}~(A), this method repeats syndrome measurements the same number of times as the code distance~$d$ and processes all the accumulated spatiotemporal syndrome information as a single batch~\cite{ueno2021dec}.
While this approach simplifies implementation, the amount of data to be processed grows as $\order{d^3}$, combining spatial~($d^2$) and temporal~($d$) dimensions, as the code distance~$d$ increases.
Moreover, when an algorithm with high time complexity such as MWPM is used, the decoding time increases rapidly with the code distance~$d$.

To meet the execution-time requirements for large code distances~$d$, sandwich decoding has been proposed~\cite{tanScalableSurfaceCode2022}.
In sandwich decoding, as shown in Fig.~\ref{fig:sandwich_batch}~(B), the graph is decomposed into multiple windows.
The window size~$w$ is defined as $w = s + 2b$ using the step size~$s$ and buffer size~$b$.
The windows are configured so that their core regions (the shaded areas in Fig.~\ref{fig:sandwich_batch}~(B)) do not overlap and collectively cover all input syndromes without gaps.

Sandwich decoding uses two types of decoders: Type-1 and Type-2.
The Type-1 decoder is a three-dimensional decoder that independently decodes each window, using the same algorithm as conventional batch decoding. The algorithm used in the Type-1 decoder is also referred to as the \emph{inner decoder}.
The Type-2 decoder is a two-dimensional decoder used to maintain consistency across core regions of adjacent windows.

The decoding procedure is as follows.
First, as in step~1 of Fig.~\ref{fig:sandwich_batch}~(B), each Type-1 decoder performs batch decoding on its respective window.
Next, as in step~2 of Fig.~\ref{fig:sandwich_batch}~(B), the error chains output by the two Type-1 decoders at the beginning and end of the core region (one round each, called the \emph{seam}) are fed into the Type-2 decoder, which corrects them to maintain consistency.
Each Type-1 decoder has two seams, referred to as the left seam and the right seam.
At its $n$-th invocation, the Type-2 decoder receives the right seam of the $n$-th window, the left seam of the $(n{+}1)$-th window, and the XOR of the syndrome information for the corresponding round.
This processing eliminates inter-window dependencies and enables parallel decoding.
Finally, the error chains from the core regions produced by the Type-1 decoders are combined with the boundary error chains produced by the Type-2 decoder to obtain the final error chain.

\begin{figure}[tb]
    \centering
    \includegraphics[width=\columnwidth]{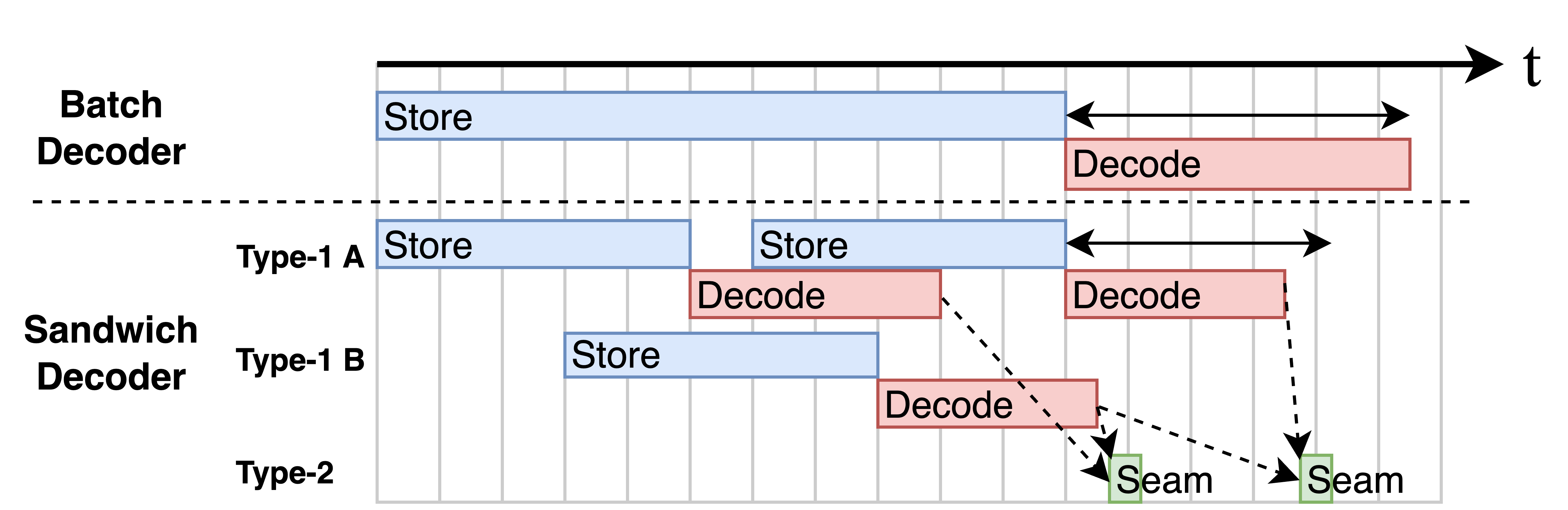}
    \caption{Difference in decode timing between sandwich decoding and batch decoding. In sandwich decoding, decoding can proceed while syndrome measurements are still being collected (Store).}
    \label{fig:timeline}
\end{figure}

By using sandwich decoding, the window size can be kept fixed even as the number of observation rounds increases, enabling decomposition of the decoding problem.
Additionally, unlike batch decoding, decoding can proceed while syndrome measurements are being collected, reducing latency.
The difference of each decoder's timeline is shown in Fig.~\ref{fig:timeline}.

However, implementing sandwich decoding in hardware presents several unique challenges.
First, because sandwich decoding shares the same syndrome information across multiple Type-1 decoders, data races must be prevented.
Second, because each Type-1 decoder completes asynchronously, the seam data supplied to the Type-2 decoder must be correctly selected from the decoding results of multiple Type-1 decoders.
Third, compared with batch decoding, the latency of the Type-2 decoder, which processes the consistency of decoding information between windows, introduces additional overhead, requiring boundary processing to be performed quickly.

\subsection{Hardware Implementations of Surface Code Decoders}

Helios, proposed by Liyanage et al., is a distributed Union-Find batch decoder implemented on multiple FPGAs~\cite{liyanage2023may}.
By using all measurement qubits as processing nodes, Helios achieves large-scale parallel processing, with latency decreasing as code distance increases. However, its computational resources scale as $\order{d^3}$ or more, making it costly.

The Collision Clustering (CC) decoder proposed by Barber et al.\ is a batch decoder based on the Union-Find decoder that introduces a memory-efficient data structure for managing error clusters and is implemented on both FPGAs and ASICs~\cite{barber2025jan}.
This decoder also proposes and uses an effective decoding method for the circuit-level noise model~\cite{dennisTopologicalQuantumMemory2002}.
However, because the CC decoder uses a data structure different from that of the standard Union-Find decoder, its compatibility with operations involving multiple logical qubits, such as lattice surgery~\cite{horsman2012dec}, remains unclear.

The QECOOL decoder proposed by Ueno et al.\ was the first decoder implemented as an online decoder that performs error decoding for each syndrome measurement~\cite{ueno2021dec}.
This decoder uses the Greedy algorithm and is implemented on single flux quantum (SFQ) circuits.
SFQ circuits are digital circuits based on superconducting technology and feature high operating speeds.

The Online Union-Find decoder proposed by Kasamura et al.\ was the first to use the Union-Find algorithm in an online decoder~\cite{kasamura2025}.
To meet the latency constraint arising from the backlog problem, this decoder limits the cluster construction step of the Union-Find decoder to one per round, thereby suppressing the processing time per round.

Sandwich decoding can be regarded as a temporally decomposed decoding approach, similar in spirit to online decoding. Its key distinction is that it coordinates and parallelizes processing across multiple rounds, offering the potential for lower decoding latency. To the best of our knowledge, no existing hardware decoder implements sandwich decoding.

\section{Design of a Sandwich Decoder Using the Union-Find Algorithm}\label{chap:proposedmethod}

\begin{figure}[tb]
    \centering
    \includegraphics[width=\columnwidth]{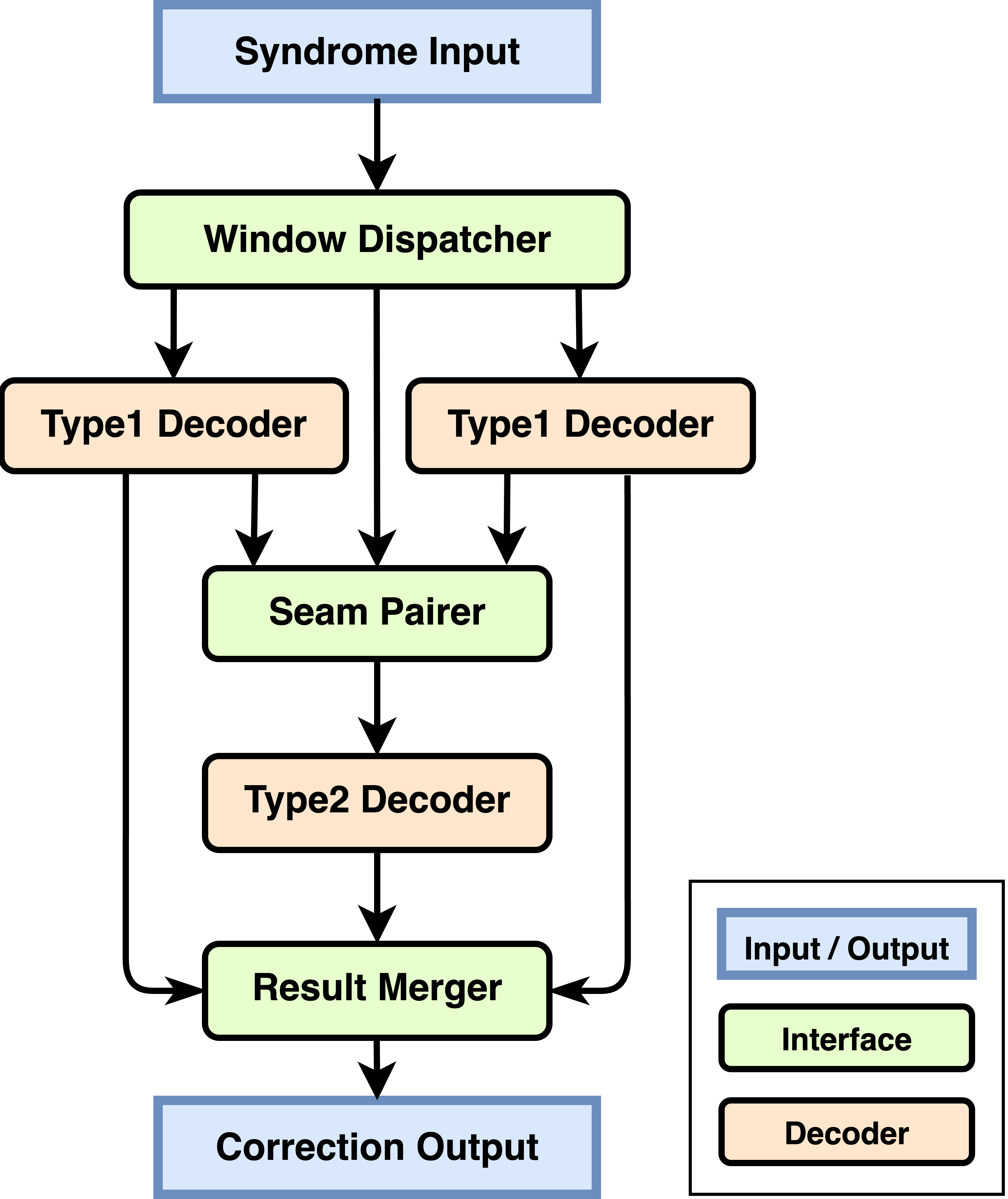}
    \caption{Microarchitecture of the sandwich decoder. It consists of five units: Window Dispatcher, Type-1 Decoder, Seam Pairer, Type-2 Decoder, and Result Merger.}
    \label{fig:microarch_block_diagram}
\end{figure}

\subsection{Proposal Overview}

In this work, we propose a parallel-processing microarchitecture for efficiently executing sandwich decoding using the Union-Find algorithm.
To address the hardware implementation challenges described above, the proposed microarchitecture includes a Window Dispatcher that manages streaming syndrome information using a ring buffer and supplies only the necessary information to the assigned Type-1 decoder.
It also includes a Seam Pairer that manages seam data using FIFOs, enabling extraction of data from the same round even when seam information is sent asynchronously, and supplies it to the Type-2 decoder.
For the Type-2 decoder, we exploit the fact that the amount of data it must process is significantly smaller than that of the Type-1 decoder to suppress the overhead of boundary processing.

In this microarchitectural design, for code distance~$d$, the step size~$s$ is set to $\lfloor(d + 1)/4\rfloor$, the buffer size~$b$ is set to $\lfloor(d - 1)/4\rfloor$, and the number of parallel Type-1 decoders is two.
The Union-Find algorithm used in this decoder is based on the hardware-optimized Union-Find algorithm proposed by Kasamura et al.\ for the Online UF decoder~\cite{kasamura2025}.

The decoder was implemented in RTL using SystemVerilog.
As shown in Fig.~\ref{fig:microarch_block_diagram}, the decoder consists of five units: Window Dispatcher, Type-1 Decoder, Seam Pairer, Type-2 Decoder, and Result Merger.
Each unit is internally controlled by a finite state machine (FSM).

\subsection{Syndrome Management by the Window Dispatcher}

In sandwich decoding, multiple Type-1 decoders require overlapping syndrome information.
The Window Dispatcher stores up to $d$ rounds of syndromes using a ring buffer and supplies the necessary syndromes to the assigned Type-1 decoder once a window's worth of data has accumulated.
Windows are alternately assigned to each Type-1 decoder.
By appropriately managing the ring buffer's read pointer, the $2b$ rounds of syndromes that overlap between consecutive windows can be reused without duplication.
Moreover, by wrapping the write pointer back to the beginning, syndrome buffering can continue, supporting the continuous decoding required in practical FTQC.

Depending on the choice of step size~$s$ and buffer size~$b$, the number of syndrome rounds supplied to the decoder at the end may be smaller than the window size $s+2b$.
This arises from the window partitioning method of sandwich decoding~\cite{tanScalableSurfaceCode2022}.
In such cases, the Window Dispatcher pads the remaining syndromes with zeros before supplying them to the Type-1 decoder.
This padding does not alter the decoding result because, for any error chain passing through the zero-padded region, a shorter path exists through the non-padded region connecting the same clusters.
Algorithm~\ref{alg:window_dispatcher} shows the implementation of the Window Dispatcher's ring queue.

\begin{algorithm}[tb]
    \caption{Ring queue of the Window Dispatcher}
    \label{alg:window_dispatcher}
    \begin{algorithmic}
        \renewcommand{\algorithmicrequire}{\textbf{Input:}}
        \renewcommand{\algorithmicensure}{\textbf{Output:}}
        \Require Streaming syndrome sequence $\{S_0, S_1, \ldots, S_{d-1}\}$
        \Ensure Window dispatch to each Type-1 decoder
        \Statex
        \Function{Dispatch}{$\textit{syndromes}$}
            \State $\textit{write\_ptr} \gets 0,\ \textit{head\_ptr} \gets 0,\ \textit{target} \gets \mathrm{A}$
            \While{not all windows dispatched}
                \If{new syndrome $S_i$ arrives} \Comment{Continuous ingestion}
                    \State $\textit{ring\_buf}[\textit{write\_ptr}] \gets S_i$
                    \State $\textit{write\_ptr} \gets (\textit{write\_ptr} + 1) \bmod d$
                \EndIf
                \If{buffered rounds $\geq s + 2b$ \textbf{and} decoder $\textit{target}$ is idle}
                    \State $\textit{read\_ptr} \gets \textit{head\_ptr}$
                    \For{$j = 0$ \textbf{to} $s + 2b - 1$}
                        \If{last window \textbf{and} $j \geq$ real rounds}
                            \State send $\mathbf{0}$ to decoder $\textit{target}$ \Comment{Padding}
                        \Else
                            \State send $\textit{ring\_buf}[\textit{read\_ptr}]$ to decoder $\textit{target}$
                            \State $\textit{read\_ptr} \gets (\textit{read\_ptr} + 1) \bmod d$
                        \EndIf
                    \EndFor
                    \State $\textit{head\_ptr} \gets (\textit{head\_ptr} + s) \bmod d$
                    \State $\textit{target} \gets \overline{\textit{target}}$ \Comment{Alternate assignment}
                \EndIf
            \EndWhile
        \EndFunction
    \end{algorithmic}
\end{algorithm}

\subsection{Seam Extraction in the Type-1 Decoder}
\begin{figure}[tb]
    \centering
    \includegraphics[width=\columnwidth]{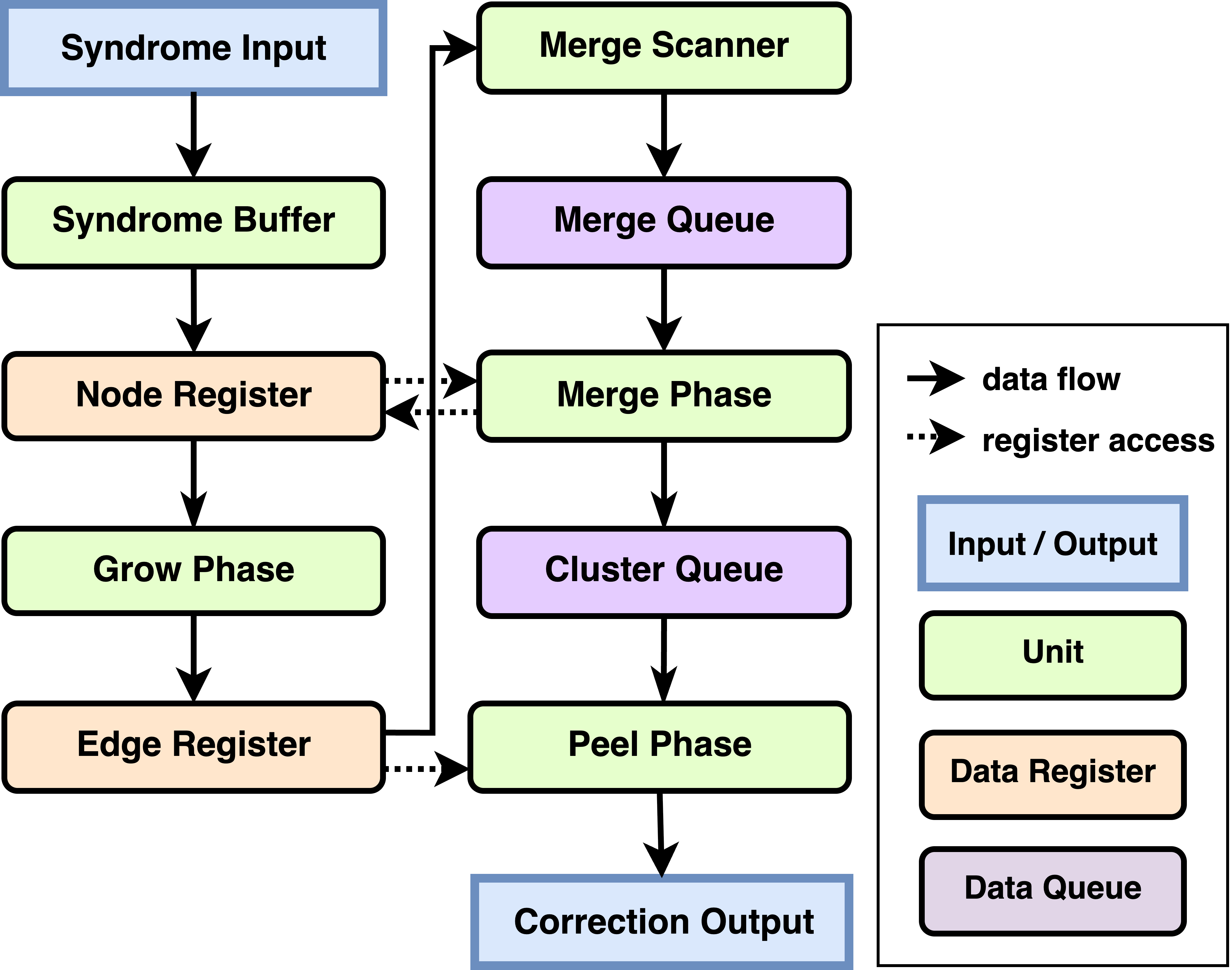}
    \caption{Internal structure of the Type-1 decoder. It consists of five units: Syndrome Buffer, Grow Phase, Merge Scanner, Merge Phase, and Peel Phase, along with registers and queues.}
    \label{fig:decoder_type1_internal}
\end{figure}

The Type-1 decoder consists of five stages as shown in Fig.~\ref{fig:decoder_type1_internal}.
Each stage is pipelined and controlled by an FSM.
This decoder is based on the hardware-optimized Union-Find algorithm proposed by Kasamura et al.~\cite{kasamura2025}, adapted for sandwich decoding.

The seam data supplied to the Type-2 decoder is extracted from the error chains generated in the Peel Phase.
Error chains are recovered by tracing edges that were involved in merging. During this process, the seam data is extracted by checking whether both endpoints of each edge in the error chain lie on a seam.
The extracted seam data is stored in a seam register.
If a seam that has already been stored is detected from another edge during seam extraction, the correct seam data can be obtained by XOR-ing the values when writing to the seam register.
Algorithm~\ref{alg:type1_peel} shows the seam extraction procedure in the Peel Phase.

The queues in Fig.~\ref{fig:decoder_type1_internal} are implemented as FIFO registers.
The depth of the Merge Queue FIFO is designed based on the maximum number of edges that a spherically grown cluster can touch within the window size.
The depth of the Cluster Queue FIFO is designed based on the expected value when errors occur independently on each edge of the three-dimensional graph with probability~$p$.

\begin{algorithm}[tb]
    \caption{Seam extraction in the Peel Phase}
    \label{alg:type1_peel}
    \begin{algorithmic}
        \renewcommand{\algorithmicrequire}{\textbf{Input:}}
        \renewcommand{\algorithmicensure}{\textbf{Output:}}
        \Require Merged edges $\mathcal{E}_{\mathrm{met}}$ from Cluster Queue, core region $[t_L, t_R]$
        \Ensure Left seam $\sigma_L$, right seam $\sigma_R$
        \Statex
        \Function{SeamSearch}{$\mathcal{E}_{\mathrm{met}}, t_L, t_R$}
            \State $\sigma_L \gets \mathbf{0},\quad \sigma_R \gets \mathbf{0}$
            \ForAll{$e_{\mathrm{met}} \in \mathcal{E}_{\mathrm{met}}$}
                \State $\textit{chain} \gets$ error chains from $e_{\mathrm{met}}$
                \ForAll{$e \in \textit{chain}$}
                    \If{$e$ is outside core region} \textbf{continue} \EndIf
                    \State $n_u, n_l \gets$ upper node, lower node of $e$
                    \If{$e$ is a temporal edge} \Comment{Temporal edge}
                        \If{$n_u.t = t_L$} $\sigma_L[n_u] \gets \sigma_L[n_u] \oplus 1$ \EndIf
                        \If{$n_l.t = t_R$} $\sigma_R[n_l] \gets \sigma_R[n_l] \oplus 1$ \EndIf
                    \Else \Comment{Spatial edge}
                        \If{$n_u.t = t_L$} $\sigma_L[n_u] \gets \sigma_L[n_u] \oplus 1$ \EndIf
                        \If{$n_l.t = t_L$} $\sigma_L[n_l] \gets \sigma_L[n_l] \oplus 1$ \EndIf
                    \EndIf
                \EndFor
            \EndFor
            \State \Return $(\sigma_L, \sigma_R)$
        \EndFunction
    \end{algorithmic}
\end{algorithm}

\subsection{Seam Data Management by the Seam Pairer}

The Seam Pairer, which connects the Type-1 decoders and the Type-2 decoder, manages the two seam outputs from each Type-1 decoder and the original syndrome information for the round corresponding to the seam position using three independent FIFOs.
Because the two Type-1 decoders alternately process windows from oldest to newest as assigned by the Window Dispatcher, the head of each FIFO always contains the oldest unprocessed seam of the respective Type-1 decoder.
This ensures that the heads of both FIFOs automatically form a matching pair for the corresponding round.
Therefore, even when the two Type-1 decoders complete asynchronously, correct seam pairing is guaranteed without the need for window-number comparison or reordering.
Algorithm~\ref{alg:seam_pairing} shows the processing of the Seam Pairer.

\begin{algorithm}[tb]
    \caption{Seam Pairing procedure}
    \label{alg:seam_pairing}
    \begin{algorithmic}
        \renewcommand{\algorithmicrequire}{\textbf{Input:}}
        \renewcommand{\algorithmicensure}{\textbf{Output:}}
        \Require Original syndrome FIFO $Q_{\mathrm{orig}}$, Type-1 seam FIFOs $Q_A, Q_B$
        \Ensure Paired seams $(O_k, R_k, L_{k+1})$ for Type-2 decoder
        \Statex
        \Function{PairSeams}{$Q_{\mathrm{orig}}, Q_A, Q_B$}
            \For{$k = 0$ \textbf{to} number of windows $- 2$}
                \If{Type-2 decoder is ready}
                \State $O_k \gets Q_{\mathrm{orig}}.\mathrm{dequeue}()$
                \If{$k$ is even}
                    \State $R_k \gets Q_A.\mathrm{front.right\_seam}$
                    \State $L_{k+1} \gets Q_B.\mathrm{front.left\_seam}$
                \Else
                    \State $R_k \gets Q_B.\mathrm{front.right\_seam}$
                    \State $L_{k+1} \gets Q_A.\mathrm{front.left\_seam}$
                \EndIf
                \State \Return $(O_k, R_k, L_{k+1})$ to Type-2 decoder
                \EndIf
            \EndFor
        \EndFunction
    \end{algorithmic}
\end{algorithm}

\subsection{Accelerating Boundary Processing in the Type-2 Decoder}

The Type-2 decoder receives the seam data from the Type-1 decoders and the syndrome information for the corresponding round as input, and outputs the error chains on the boundary between windows.
The Type-2 decoder is implemented using a two-dimensional Union-Find algorithm.
While the Type-2 decoder processes $d^2$ edges, the Type-1 decoder must process $\order{d^3}$ edges.
By exploiting the significantly smaller amount of data to be processed compared with the Type-1 decoder, we use a priority encoder to check all edges simultaneously, thereby accelerating boundary processing.

The queues within the Type-2 decoder are implemented as FIFO registers, similar to the Type-1 decoder, but the FIFOs are designed to be shallower due to the difference in edge count.
The Merge Queue is designed based on the maximum number of edges that a circularly grown cluster can touch, taking advantage of the fact that the Type-2 decoder is two-dimensional.
The Cluster Queue is designed based on the expected value when errors on each edge of the two-dimensional graph occur independently with probability~$p$.

\section{Evaluation}\label{chap:evaluation}

\subsection{Evaluation Environment}

For FTQC, practical requirements such as implementing thousands of logical qubits in a cryogenic environment~\cite{charbon2016feb} make speed (latency) and area the two primary metrics for decoders.
In this work, we perform a comparative evaluation of the proposed Union-Find sandwich decoder against a conventional Union-Find batch decoder using these metrics.
Hereafter, including in tables, the two are referred to simply as the sandwich decoder (Sandw.) and the batch decoder (Batch).
Both decoders share the same Union-Find algorithm microarchitecture as presented in the proposed method.

A Rust-based simulator was used for threshold evaluation.
RTL simulations using Verilator were performed for cycle-count evaluation.
Error generation for both evaluations used Stim~\cite{gidney2021stim}.
ASIC implementation was performed through logic synthesis using Synopsys Design Compiler, targeting TSMC CMOS \SI{22}{\nano\meter} and ASAP FinFET \SI{7}{\nano\meter}~\cite{CLARK2016105}.

\begin{figure*}[tb]
    \centering
    \begin{minipage}[b]{0.32\textwidth}
        \centering
        \includegraphics[width=\textwidth]{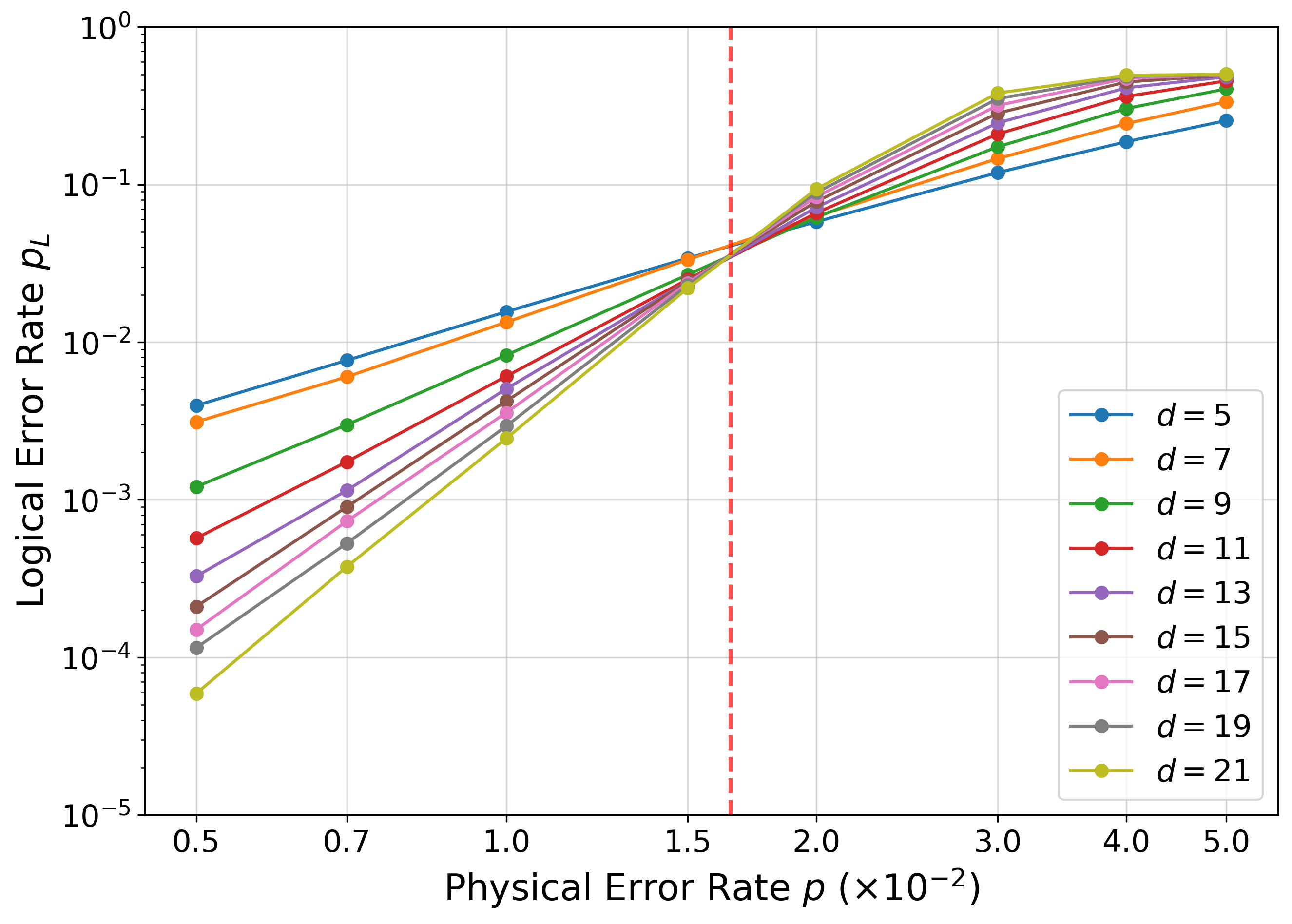}
        (a) Batch UF (phenomenological).
    \end{minipage}
    \begin{minipage}[b]{0.32\textwidth}
        \centering
        \includegraphics[width=\textwidth]{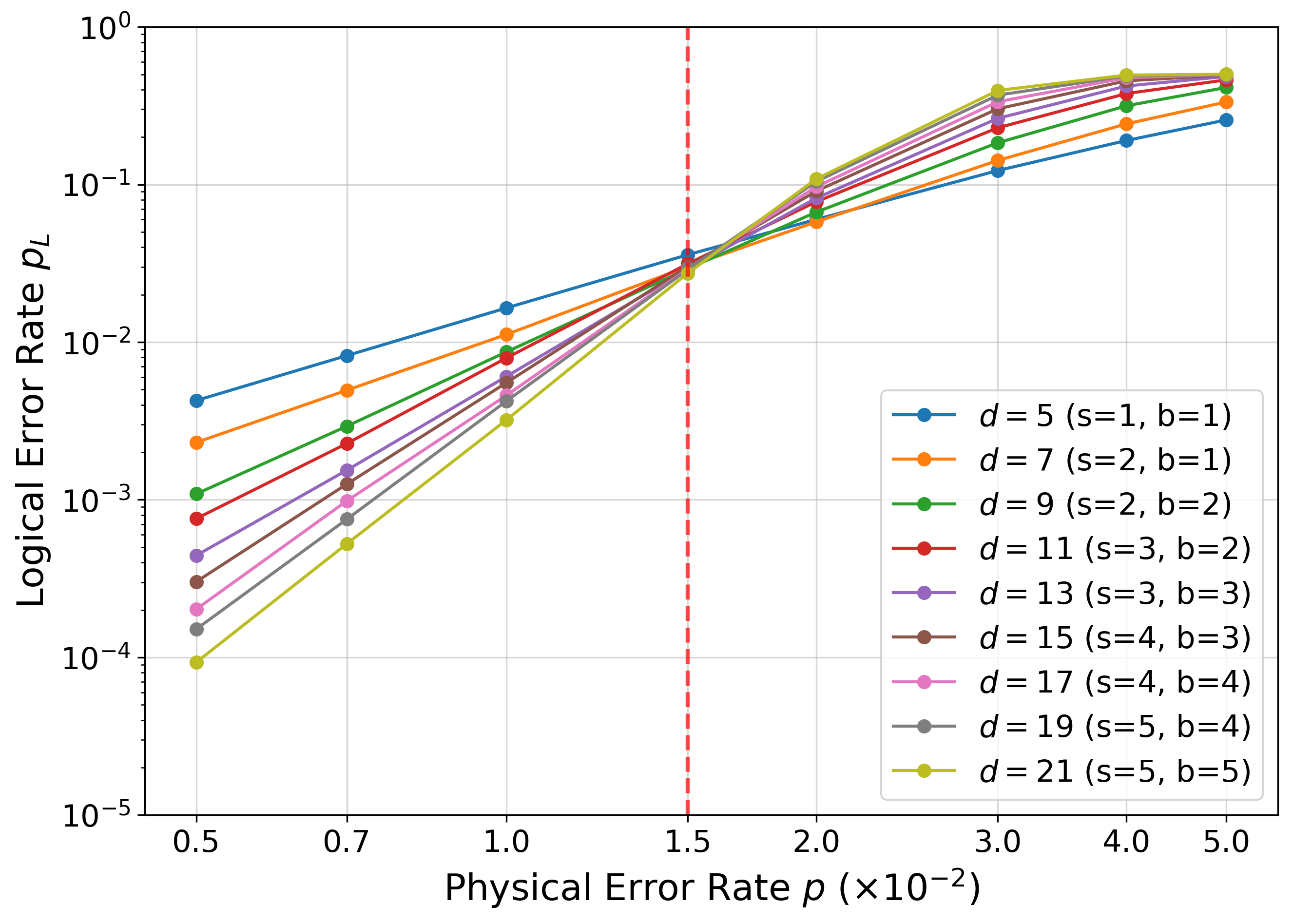}
        (b) Sandwich UF (phenomenological).
    \end{minipage}
    \begin{minipage}[b]{0.32\textwidth}
        \centering
        \includegraphics[width=\textwidth]{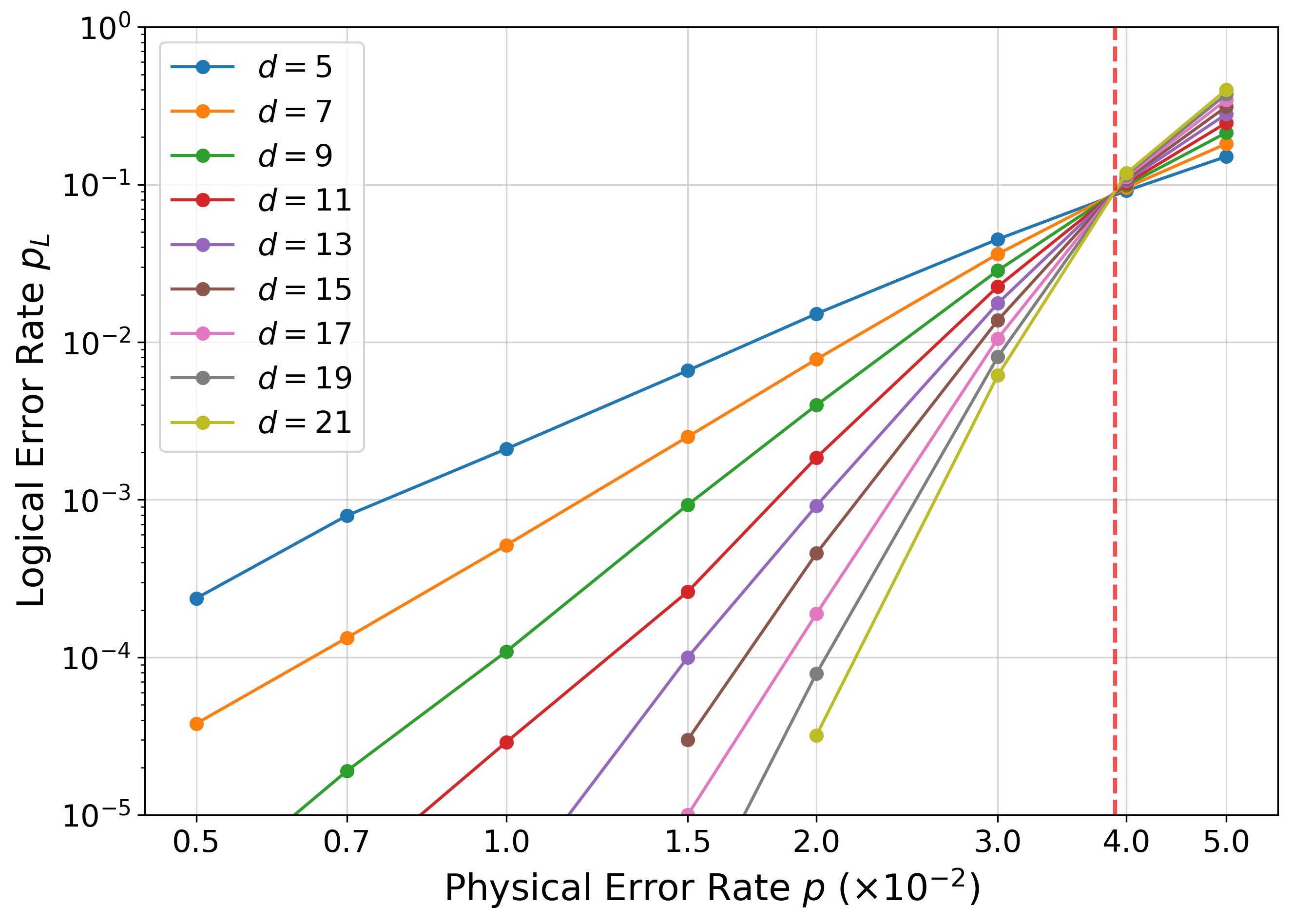}
        (c) PyMatching (phenomenological).
    \end{minipage}

    \vspace{0.5em}

    \begin{minipage}[b]{0.32\textwidth}
        \centering
        \includegraphics[width=\textwidth]{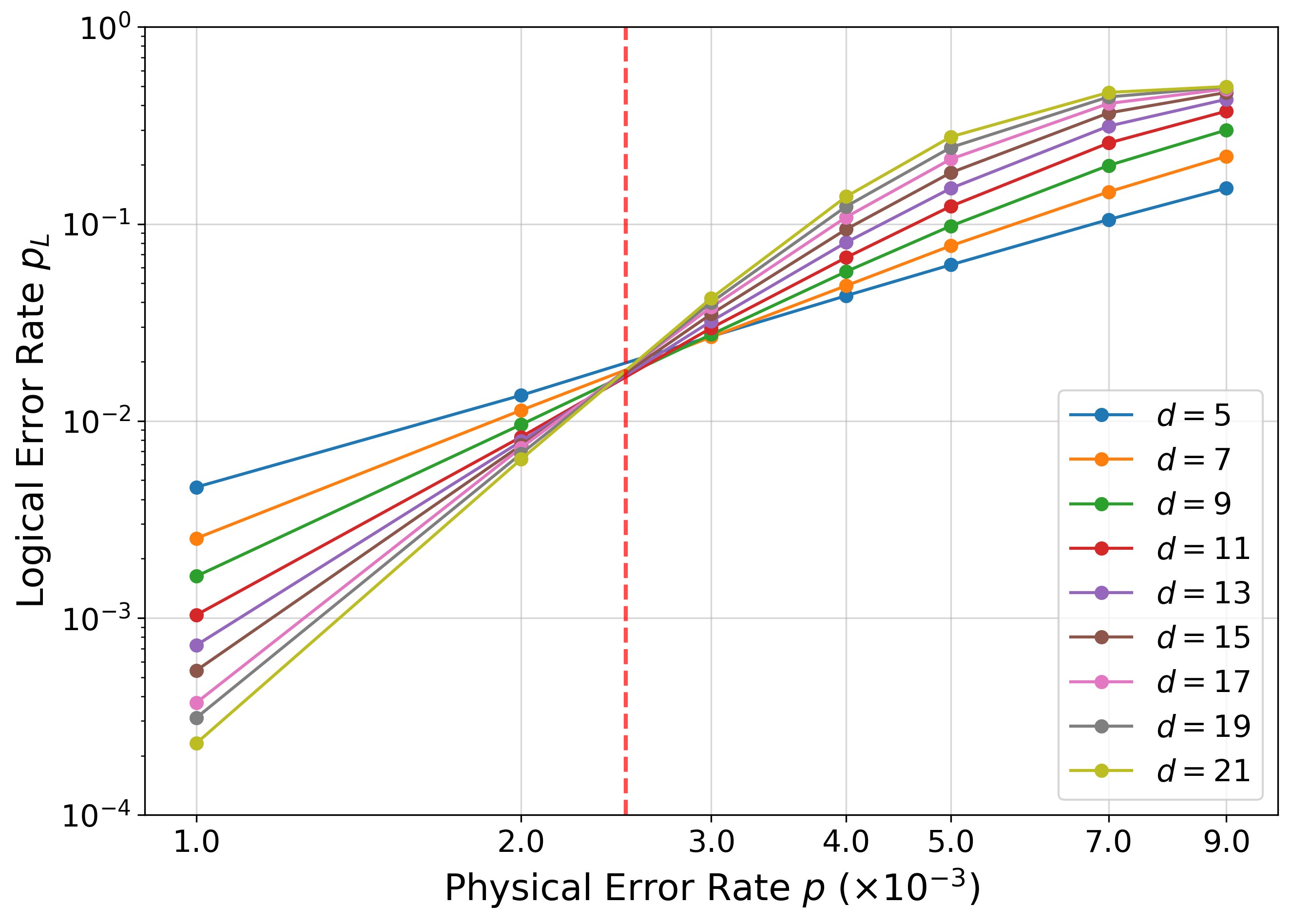}
        (d) Batch UF (circuit-level).
    \end{minipage}
    \begin{minipage}[b]{0.32\textwidth}
        \centering
        \includegraphics[width=\textwidth]{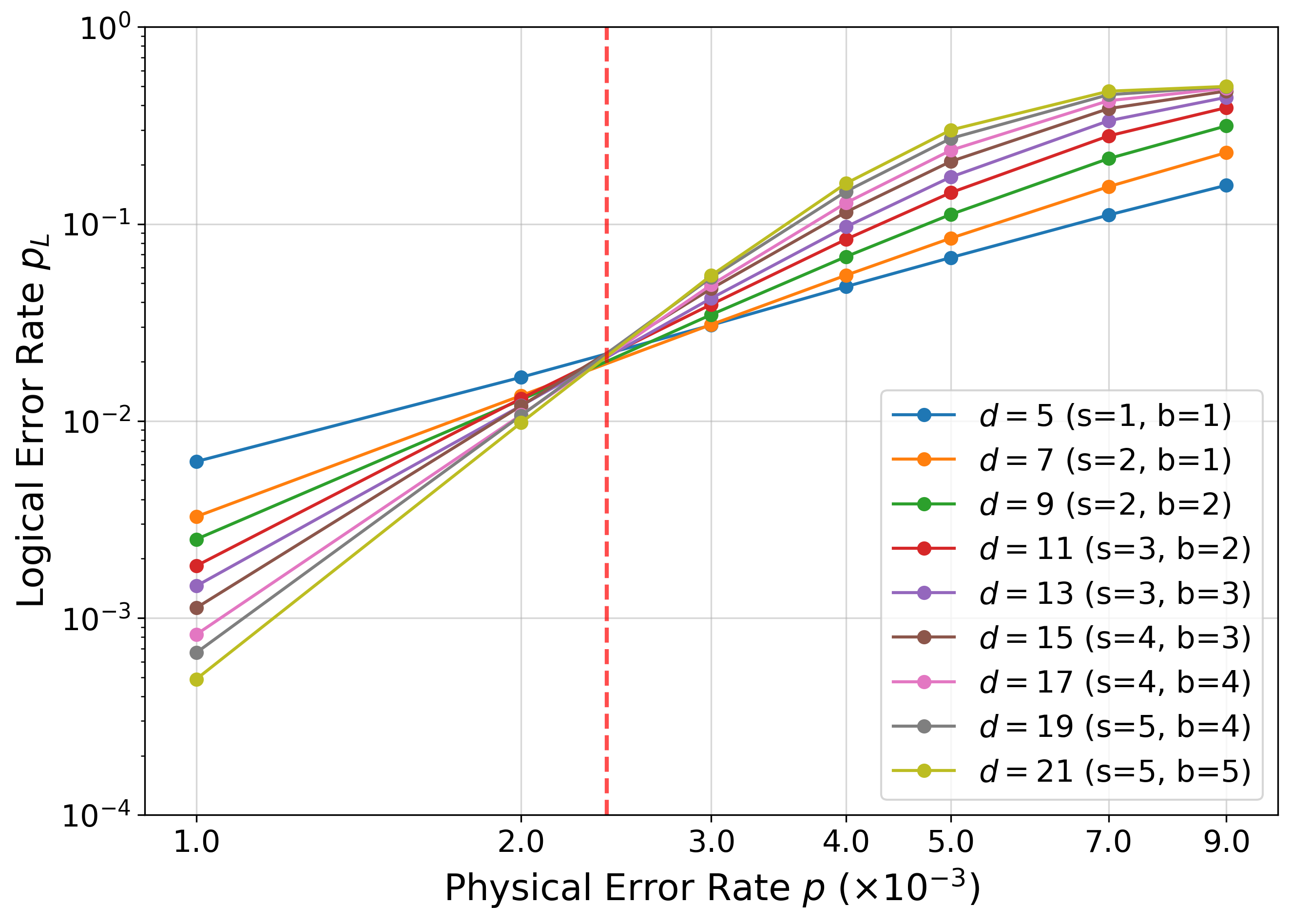}
        (e) Sandwich UF (circuit-level).
    \end{minipage}
    \begin{minipage}[b]{0.32\textwidth}
        \centering
        \includegraphics[width=\textwidth]{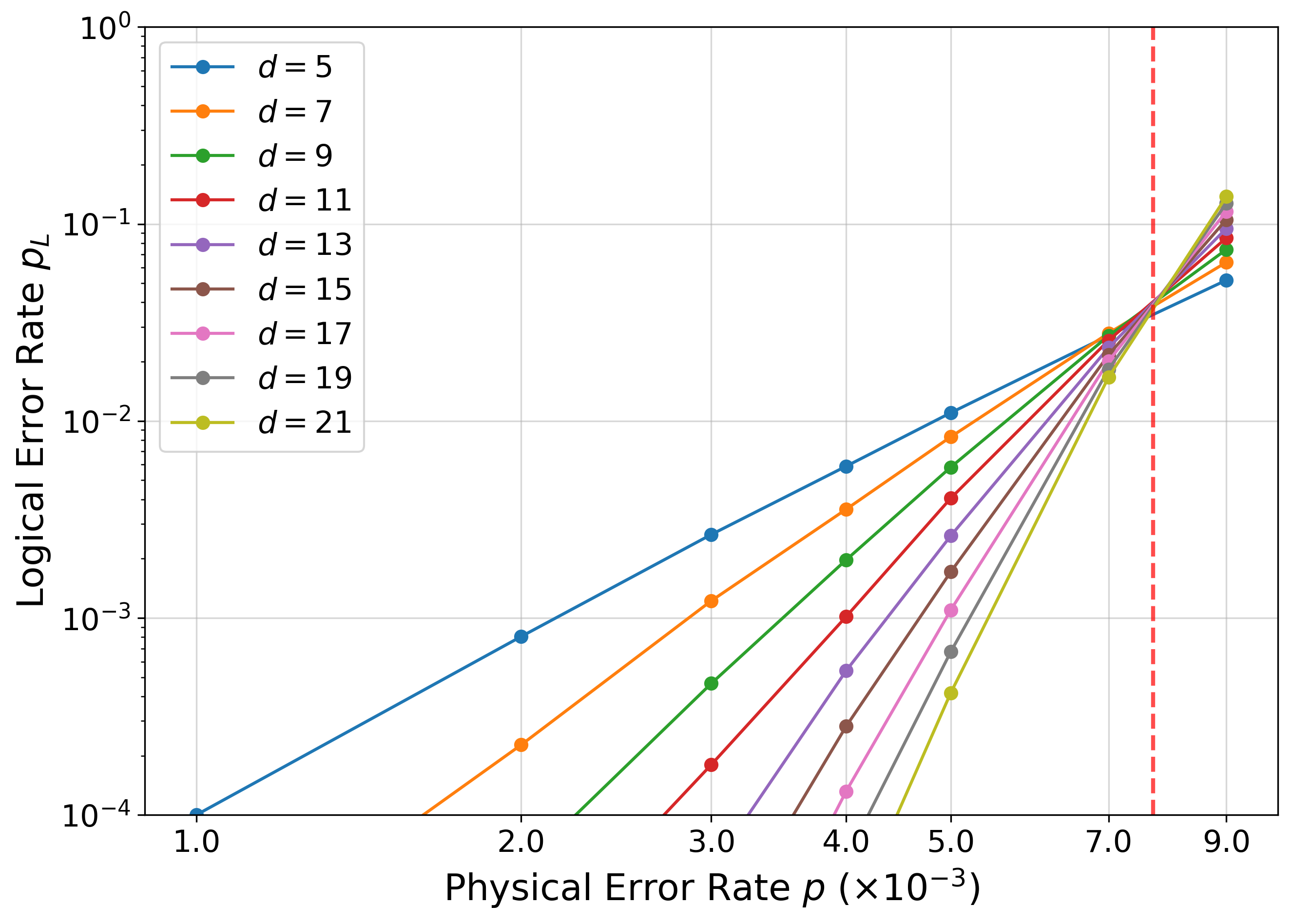}
        (f) PyMatching (circuit-level).
    \end{minipage}

    \caption{Threshold evaluation under (top row) phenomenological noise and (bottom row) Stim's standard circuit-level noise. PyMatching\cite{higgott2022pymatching} is run on identical Stim-generated samples as a MWPM baseline. The red dashed line indicates the threshold of each decoder. Points are generated using a maximum of $10^6$ samples and the points with no more than 1 error are removed.}
    \label{fig:threshold}
\end{figure*}

\subsection{Threshold Evaluation}

Threshold evaluation of the decoder is performed by measuring the probability~$P_L$ that the decoding result causes a logical error after $d$ rounds of syndrome measurements for a code of distance~$d$~\cite{liyanage2023may, kasamura2025, barber2025jan}.
This evaluation was performed for $d = 5, 7, 9, 11, 13, 15, 17, 19, 21$, with $10^6$ Monte Carlo simulations for each data point.
The simulator was implemented in Rust, and for each code distance, the outputs were compared with those of the RTL for $10^4$ runs to confirm agreement.

We evaluated under two noise models. The first is a phenomenological model with $(p_I, p_X, p_Y, p_Z) = (1-p, p/3, p/3, p/3)$ on data qubits and an independent measurement-flip probability~$p$, generated with Stim by setting the \path{before_measure_flip_probability} and \path{before_round_data_depolarization} parameters to~$p$ and leaving all others at zero. Since decoding for $X$ and $Z$ errors is independent, only $X$-error decoding is evaluated.
The second is a circuit-level noise model, in which a uniform error rate~$p$ is applied to every gate, idle, reset, and measurement operation in the syndrome extraction circuit. This noise model was generated by setting all the Stim's parameters to $p$.
For each noise model, we additionally run PyMatching~\cite{higgott2022pymatching} on the same Stim-generated error samples as a MWPM baseline. We focus primarily on phenomenological noise since it enables direct comparison with prior decoder-algorithm work~\cite{ueno2021dec, kasamura2025}. We additionally report circuit-level noise model results as a more realistic evaluation. 

Fig.~\ref{fig:threshold} shows the threshold evaluation results.
Under phenomenological noise, the threshold is \SI{1.6}{\percent} for the batch UF decoder, \SI{1.5}{\percent} for the sandwich UF decoder, and approximately \SI{3.5}{\percent} for PyMatching.
Under circuit-level noise, the batch UF and sandwich UF decoders both achieve a threshold of approximately \SI{0.25}{\percent}, while PyMatching achieves approximately \SI{0.78}{\percent}.
Under both noise models, the proposed sandwich decoder exhibits only a slight degradation relative to the batch UF decoder, while a gap to PyMatching is observed. This gap reflects the accuracy difference between approximate Union-Find and exact MWPM~\cite{delfosse2021dec}, as our Union-Find inner decoder treats all detector-graph edges with uniform weight, whereas MWPM exploits the heterogeneous edge weights derived from Stim's detector error model.

In comparison with prior hardware decoders evaluated under the same phenomenological noise model, the Greedy-based hardware decoder QECOOL~\cite{ueno2021dec} has been reported at \SI{2}{\percent}~\cite{uenoQULATIS2022}, and the hardware online Union-Find decoder~\cite{kasamura2025} at \SI{1.4}{\percent}. The threshold of \SI{1.5}{\percent} achieved by the proposed sandwich decoder is comparable to these existing hardware decoders.

\subsection{Cycle Count Evaluation}

For cycle count evaluation, $d$ rounds of syndrome measurements were performed, and the number of cycles required for the batch decoder and sandwich decoder to complete decoding was measured.
Cycle counts were evaluated through RTL simulation using Verilator.
Noise was generated using Stim as in the threshold evaluation, with simulations conducted with the phenomenological noise model under $p = \SI{1.0}{\percent}$ and $p = \SI{0.1}{\percent}$.
Each data point was evaluated over $10^4$ simulations.
The syndrome input period was set to one syndrome input every 1000 cycles, assuming the \SI{1}{\micro\second} syndrome measurement cycle of superconducting qubits, so that the backlog problem does not occur~\cite{byunXQsimModelingCrosstechnology2022}.
Table~\ref{tab:execution_cycles} shows the cycle count evaluation results for each code distance~$d$.
For each data point, the maximum, average, and standard deviation ($\sigma$) of the number of decode cycles divided by the number of measurement rounds are shown.

\begin{table}[tb]
    \centering
    \renewcommand{\arraystretch}{0.6}
    \caption{Cycle count evaluation results per round $(n = 10^4)$}
    \label{tab:execution_cycles}
    \resizebox{\columnwidth}{!}{
    \begin{tabular}{r|c| ccc ccc}
        \toprule
        & &\multicolumn{3}{c}{$p = \SI{1.0}{\percent}$} & \multicolumn{3}{c}{$p = \SI{0.1}{\percent}$} \\
        \cmidrule(lr){3-5} \cmidrule(lr){6-8}
        $d$                 & Method & Avg.  & Max.   & $\sigma$  & Avg. & Max.  & $\sigma$  \\
        \midrule
        \multirow{2}{*}{5}  & Batch & 8.9   & 112.0  & 9.1       & 2.3  & 73.2  & 2.2 \\
                            & Sandw. & 9.8   & 93.6   & 7.7       & 5.1  & 43.2  & 1.8 \\
        \midrule
        \multirow{2}{*}{7}  & Batch & 18.6  & 239.9  & 20.7      & 2.7  & 137.9 & 3.5 \\
                            & Sandw. & 12.9  & 160.6  & 14.4      & 4.2  & 147.4 & 2.6 \\
        \midrule
        \multirow{2}{*}{9}  & Batch & 33.8  & 287.8  & 32.0      & 3.8  & 219.2 & 4.7 \\
                            & Sandw. & 24.1  & 263.6  & 25.4      & 4.7  & 110.9 & 3.5 \\
        \midrule
        \multirow{2}{*}{11} & Batch & 55.4  & 442.0  & 43.7      & 5.7  & 188.9 & 5.7 \\
                            & Sandw. & 32.1  & 440.2  & 31.1      & 5.0  & 125.0 & 3.8 \\
        \midrule
        \multirow{2}{*}{13} & Batch & 85.9  & 728.5  & 60.6      & 8.2  & 163.1 & 6.3 \\
                            & Sandw. & 54.7  & 557.9  & 44.9      & 6.8  & 198.2 & 5.3 \\
        \midrule
        \multirow{2}{*}{15} & Batch & 122.3 & 961.9  & 70.8      & 11.2 & 181.9 & 7.8 \\
                            & Sandw. & 69.2  & 740.5  & 54.4      & 7.7  & 181.1 & 6.3 \\
        \midrule
        \multirow{2}{*}{17} & Batch & 171.0 & 1249.6 & 91.9      & 14.9 & 168.7 & 8.2 \\
                            & Sandw. & 107.5 & 1066.9 & 70.8      & 10.6 & 158.1 & 6.7 \\
        \midrule
        \multirow{2}{*}{19} & Batch & 232.1 & 1413.8 & 111.4     & 19.4 & 205.3 & 9.5 \\
                            & Sandw. & 137.2 & 1204.1 & 91.2      & 12.2 & 201.3 & 7.6 \\
        \midrule
        \multirow{2}{*}{21} & Batch & 301.3 & 1902.6 & 134.1     & 24.9 & 200.1 & 11.0\\
                            & Sandw. & 196.9 & 1683.7 & 120.8     & 16.3 & 214.6 & 9.2 \\
        \midrule
        \multirow{2}{*}{23} & Batch & 385.8 & 2523.1 & 167.2     & 31.2 & 191.6 & 11.8\\
                            & Sandw. & 244.7 & 1816.0 & 143.3     & 18.4 & 183.5 & 9.2 \\
        \bottomrule
    \end{tabular}
    }
\end{table}

\begin{table*}[tb]
    \centering
    \caption{Maximum frequency, circuit area, and decode latency evaluation by logic synthesis for each code distance~$d$}
    \label{tab:asic_evaluation}

    \vspace{2mm}
    {\small (a) TSMC CMOS \SI{22}{\nano\meter}} \\
    \renewcommand{\arraystretch}{0.75}
    \setlength{\tabcolsep}{4pt}
    \resizebox{\textwidth}{!}{
        \begin{tabular}{r|cc cc cc cc cc cc}
            \toprule\toprule

            Code dist.
            & \multicolumn{2}{c}{\makecell{Max. freq.\\{} [MHz]}}
            & \multicolumn{2}{c}{\makecell{Circuit area\\{} [mm$^2$]}}
            & \multicolumn{2}{c}{\makecell{Latency (avg.)\\($p=\SI{1.0}{\percent}$)[ns] }}
            & \multicolumn{2}{c}{\makecell{Latency (avg.)\\($p=\SI{0.1}{\percent}$)[ns] }}
            & \multicolumn{2}{c}{\makecell{Latency (max.)\\($p=\SI{1.0}{\percent}$)[ns] }}
            & \multicolumn{2}{c}{\makecell{Latency (max.)\\($p=\SI{0.1}{\percent}$)[ns] }} \\

            \cmidrule(lr){2-3} \cmidrule(lr){4-5} \cmidrule(lr){6-7} \cmidrule(lr){8-9} \cmidrule(lr){10-11} \cmidrule(lr){12-13}
            $d$ & Batch & Sandw. & Batch   & Sandw.   & Batch & Sandw. & Batch  & Sandw.  & Batch & Sandw.  & Batch & Sandw.   \\
            \midrule
            5   & 2703 & 2703 & 0.0343 & 0.0543 & 3.3  & 3.6  & 0.9  & 1.9   & 41.4  & 34.6  & 27.1 & 16.0   \\
            7   & 2273 & 2326 & 0.0815 & 0.1163 & 8.2  & 5.6  & 1.2  & 1.8   & 105.5 & 69.0  & 60.6 & 63.4   \\
            9   & 2041 & 2083 & 0.1983 & 0.2789 & 16.6 & 11.6 & 1.9  & 2.3   & 141.0 & 126.5 & 107.4& 53.2   \\
            11  & 2000 & 2000 & 0.3836 & 0.4714 & 27.7 & 16.0 & 2.8  & 2.5   & 221.0 & 220.1 & 94.5 & 62.5   \\
            13  & 1923 & 1923 & 0.5464 & 0.8449 & 44.7 & 28.4 & 4.2  & 3.5   & 378.9 & 290.1 & 84.8 & 103.0  \\
            15  & 1887 & 1754 & 0.8948 & 1.2002 & 64.8 & 39.4 & 5.9  & 4.4   & 510.0 & 422.2 & 96.4 & 103.3  \\
            17  & 1724 & 1695 & 1.3569 & 2.0487 & 99.2 & 63.4 & 8.7  & 6.3   & 724.8 & 629.5 & 97.9 & 93.3   \\
            19  & 1639 & 1587 & 1.9245 & 2.6685 & 141.6& 86.4 & 11.9 & 7.7   & 862.6 & 758.7 & 125.3& 126.9  \\
            21  & 1613 & 1613 & 2.5444 & 3.5096 & 186.7& 122.1& 15.4 & 10.1  & 1179.5& 1043.8& 124.1& 133.1  \\
            23  & --   & 1587 &  --    & 5.1166 &  --  & 154.2&  --  & 11.6  &  --   & 1144.3&  --  & 115.6  \\
            \bottomrule
        \end{tabular}
    }

    \vspace{1.5em}

    {\small (b) ASAP FinFET \SI{7}{\nano\meter}} \\
    \renewcommand{\arraystretch}{0.75}
    \setlength{\tabcolsep}{4pt}
    \resizebox{\textwidth}{!}{
        \begin{tabular}{r|cc cc cc cc cc cc}
            \toprule\toprule
            Code dist.
            & \multicolumn{2}{c}{\makecell{Max. freq.\\{} [MHz]}}
            & \multicolumn{2}{c}{\makecell{Circuit area\\{} [mm$^2$]}}
            & \multicolumn{2}{c}{\makecell{Latency (avg.)\\($p=\SI{1.0}{\percent}$)[ns] }}
            & \multicolumn{2}{c}{\makecell{Latency (avg.)\\($p=\SI{0.1}{\percent}$)[ns] }}
            & \multicolumn{2}{c}{\makecell{Latency (max.)\\($p=\SI{1.0}{\percent}$)[ns] }}
            & \multicolumn{2}{c}{\makecell{Latency (max.)\\($p=\SI{0.1}{\percent}$)[ns] }} \\

            \cmidrule(lr){2-3} \cmidrule(lr){4-5} \cmidrule(lr){6-7} \cmidrule(lr){8-9} \cmidrule(lr){10-11} \cmidrule(lr){12-13}
            $d$ & Batch & Sandw. & Batch   & Sandw.   & Batch & Sandw. & Batch & Sandw. & Batch  & Sandw.  & Batch & Sandw.  \\
            \midrule
            5   & 3121 & 3175 & 0.0073 & 0.0115 & 2.8  & 3.1  & 0.7  & 1.6  & 35.9  & 29.5  & 23.5 & 13.6  \\
            7   & 2940 & 3003 & 0.0191 & 0.0274 & 6.3  & 4.3  & 0.9  & 1.4  & 81.6  & 53.5  & 46.9 & 49.1  \\
            9   & 2705 & 2774 & 0.0452 & 0.0673 & 12.5 & 8.7  & 1.4  & 1.7  & 106.4 & 95.0  & 81.0 & 40.0  \\
            11  & 2564 & 2564 & 0.0873 & 0.1072 & 21.6 & 12.5 & 2.2  & 2.0  & 172.4 & 171.7 & 73.7 & 48.8  \\
            13  & 2326 & 2325 & 0.1285 & 0.1935 & 36.9 & 23.5 & 3.5  & 2.9  & 313.2 & 240.0 & 70.1 & 85.2  \\
            15  & 2123 & 2179 & 0.2084 & 0.2777 & 57.6 & 31.7 & 5.3  & 3.6  & 453.1 & 339.9 & 85.7 & 83.1  \\
            17  & 2128 & 2128 & 0.3214 & 0.4611 & 80.4 & 50.5 & 7.0  & 5.0  & 587.2 & 501.4 & 79.3 & 74.3  \\
            19  & 2021 & 2041 & 0.4445 & 0.6228 & 114.8& 67.2 & 9.6  & 6.0  & 699.5 & 589.9 & 101.6& 98.6  \\
            21  & 1961 & 1961 & 0.6119 & 0.8309 & 153.6& 100.4& 12.7 & 8.3  & 970.2 & 858.6 & 102.1& 109.4 \\
            23  & --   & 1923 & --     & 1.1875 & --   & 127.2& --   & 9.6  & --    & 944.4 & --   & 95.4  \\
            \bottomrule
        \end{tabular}
    }
\end{table*}

\subsection{ASIC implementation}

For the ASIC implementation, we performed logic synthesis for both the batch decoder and the sandwich decoder assuming $d$ rounds of syndrome measurements. We evaluated the area, maximum operating frequency, and maximum/average latency. Latency values are derived from the cycle count results under the phenomenological noise model at $p = \SI{1.0}{\percent}$ and $p = \SI{0.1}{\percent}$.
Logic synthesis was performed using Synopsys Design Compiler, with implementation on two processes: TSMC CMOS \SI{22}{\nano\meter} and ASAP FinFET \SI{7}{\nano\meter}~\cite{CLARK2016105}.
Latency was computed from the cycle count evaluation results and the maximum operating frequency.
Logic synthesis of the batch decoder at $d=23$ failed to complete due to virtual memory exhaustion on our synthesis host (process limit set to \SI{450}{\giga\byte}), so the corresponding entries in Table~\ref{tab:asic_evaluation} are left blank.

\begin{figure}[tb]
    \centering
    \includegraphics[width=0.54\columnwidth]{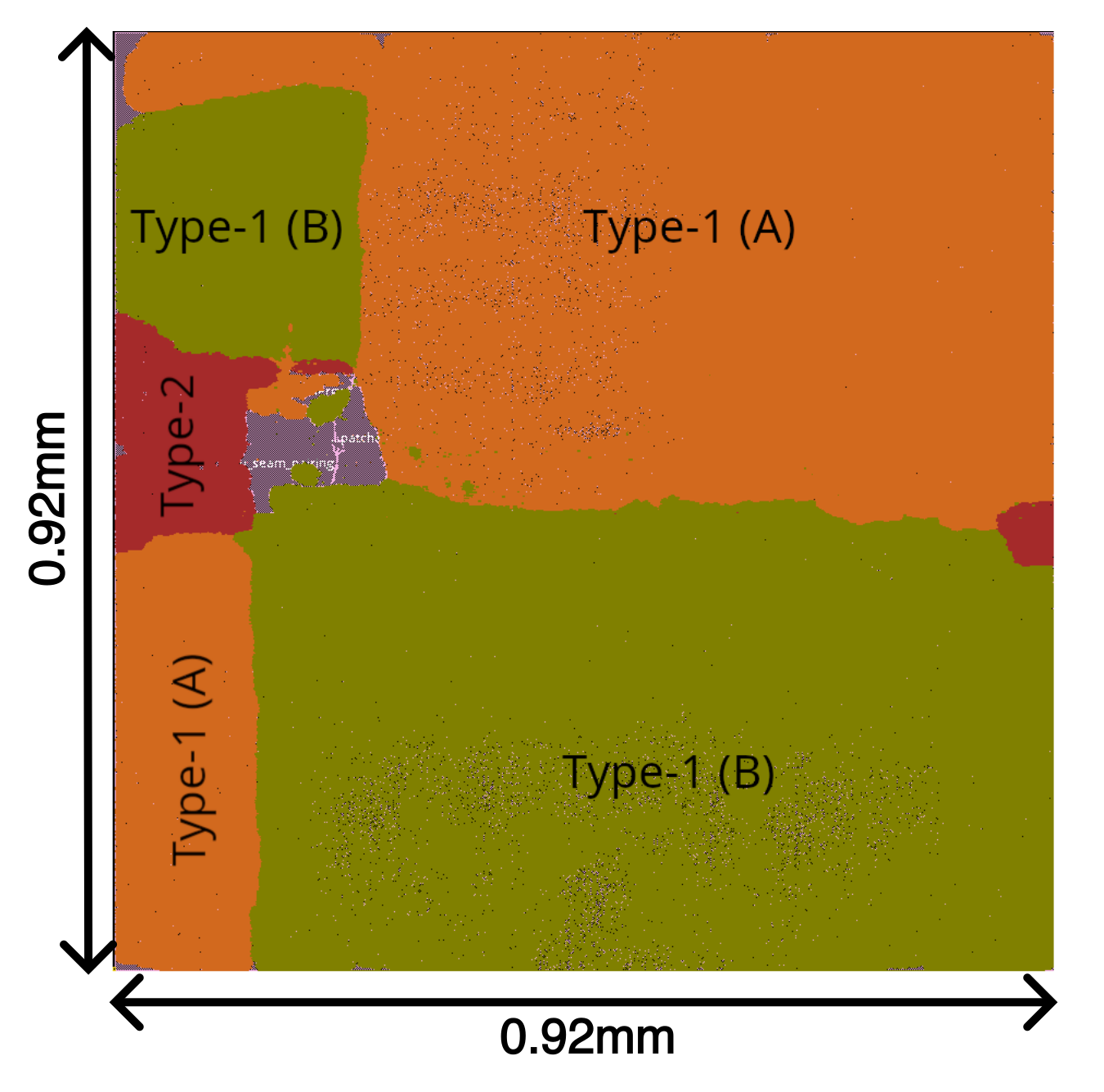}
    \caption{Place-and-route layout of the proposed sandwich decoder at $d = 13$ on the \SI{22}{\nano\meter} process.}
    \label{fig:innovus_layout}
\end{figure}

Table~\ref{tab:asic_evaluation} shows the ASIC implementation results.
The proposed sandwich decoder achieves a \SI{34.9}{\percent} reduction in average latency and an \SI{11.5}{\percent} reduction in worst-case latency at physical error rate $p = \SI{1.0}{\percent}$ and code distance $d = 21$ on both the \SI{7}{\nano\meter} and \SI{22}{\nano\meter} processes compared with the batch decoder.
At physical error rate $p = \SI{0.1}{\percent}$, the sandwich decoder shows a \SI{34.6}{\percent} reduction in average latency at $d = 21$, while worst-case latency increases by \SI{7.2}{\percent}.
On the other hand, circuit area increases by a factor of 1.3 to 1.5 for the sandwich decoder compared with the batch decoder.
For all code distances at which synthesis completed, both decoders achieve average latency below \SI{1}{\micro\second}, demonstrating that high-speed decoding is feasible.
Fig.~\ref{fig:innovus_layout} shows the place-and-route layout of the proposed decoder at $d = 13$ on the \SI{22}{\nano\meter} process for reference.

\begin{figure}[tb]
    \centering
    \begin{minipage}[b]{0.49\columnwidth}
        \centering
        \includegraphics[width=\textwidth]{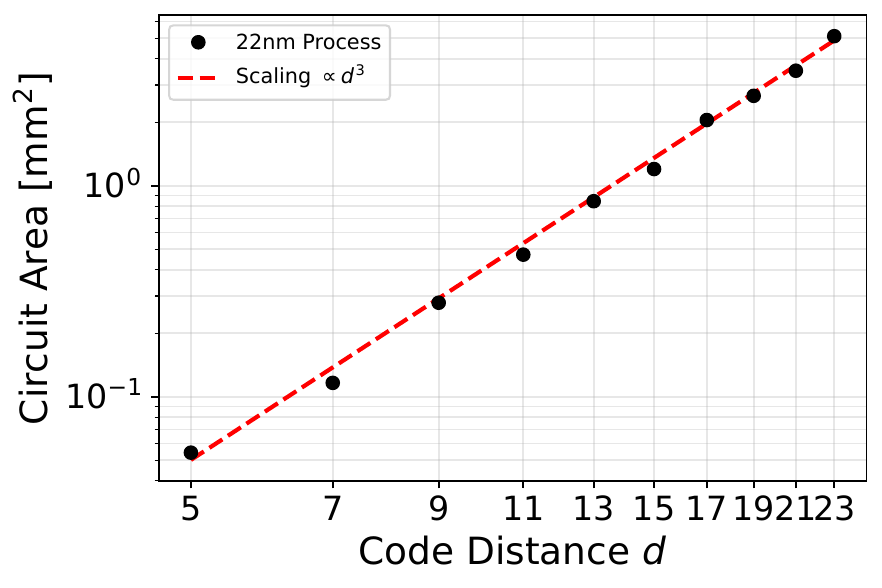}
        (a) \SI{22}{\nano\meter} process.
    \end{minipage}\hfill
    \begin{minipage}[b]{0.49\columnwidth}
        \centering
        \includegraphics[width=\textwidth]{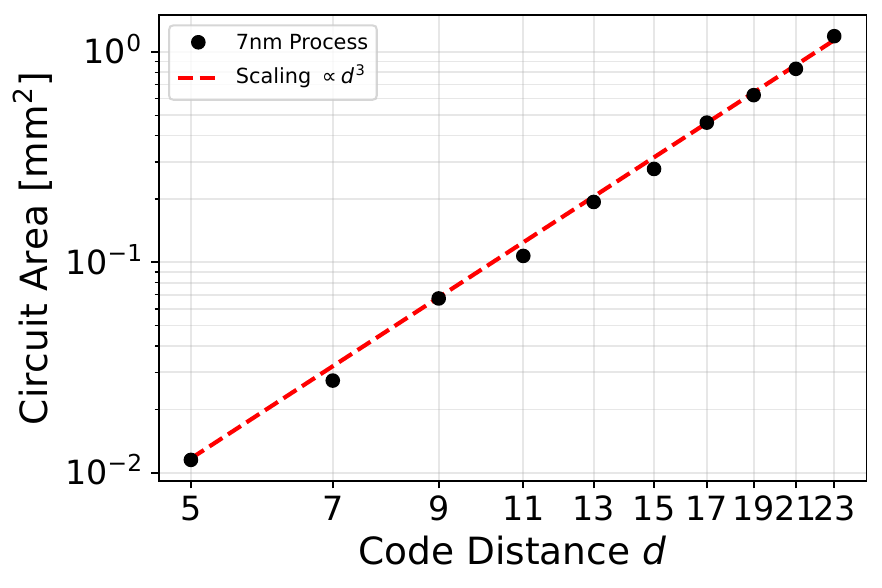}
        (b) \SI{7}{\nano\meter} process.
    \end{minipage}
    \caption{Scaling of the circuit area of the proposed sandwich decoder at (a) \SI{22}{\nano\meter} and (b) \SI{7}{\nano\meter} processes. The area scales as $\order{d^3}$.}
    \label{fig:scaling}
\end{figure}

As shown in Fig.~\ref{fig:scaling}, the circuit area of the sandwich decoder scales as $\order{d^3}$ with respect to the code distance~$d$.
This is because the registers representing the graph, which dominate the Union-Find algorithm microarchitecture, require $\order{d^3}$ resources.
The sandwich decoder has two such graph registers, each storing $\lfloor(d+1)/4\rfloor + 2 \times \lfloor(d-1)/4\rfloor$ rounds of graph information.
In contrast, the batch decoder has one graph register storing $d$ rounds of graph information.
The ratio of these two round counts can be evaluated as:
\begin{align}
    \frac{2 \times(\lfloor(d + 1) /4 \rfloor+ 2 \times \lfloor(d - 1) / 4\rfloor)}{d} & \leq \frac{3}{2} - \frac{1}{2d}
\end{align}
This shows that the sandwich decoder requires at most 1.5 times as many graph registers as the batch decoder, which contributes to the increase in circuit area.

\begin{figure}[tb]
    \centering
    \includegraphics[width=0.9\columnwidth]{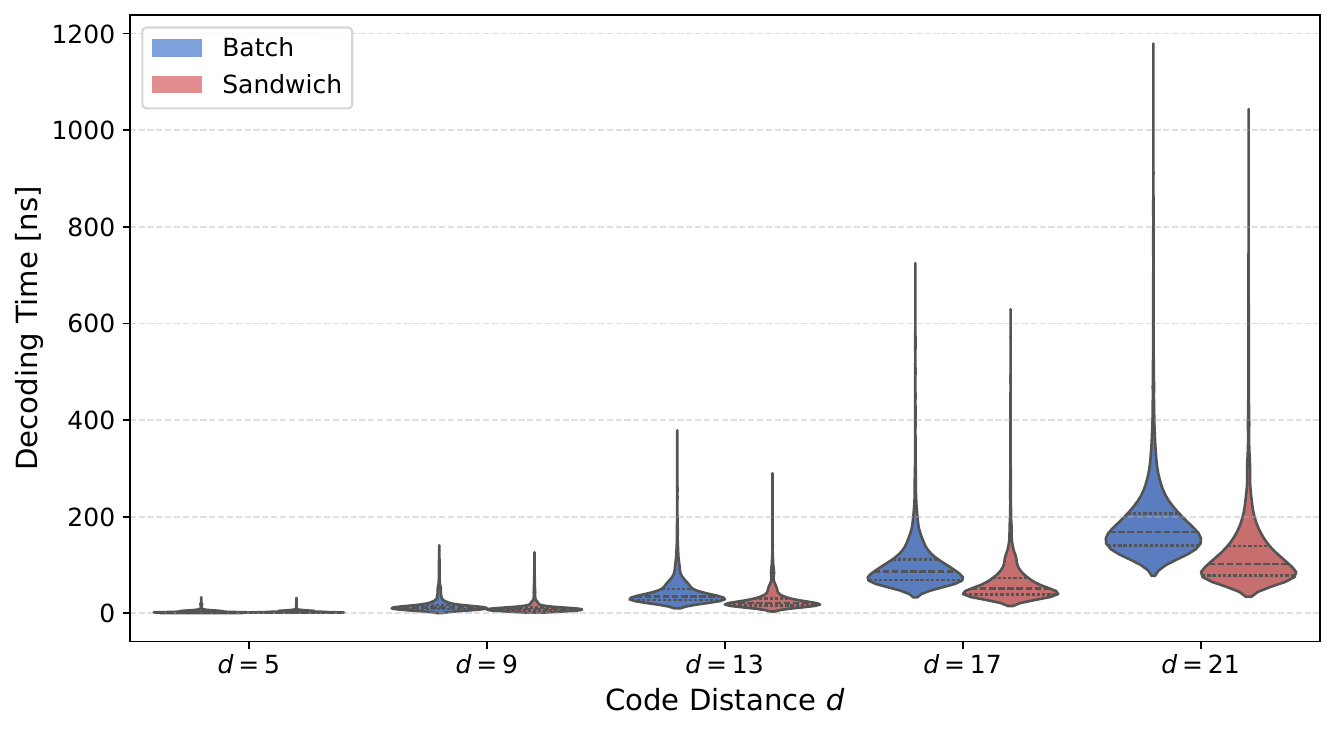}
    \caption{Decode latency distribution for code distance~$d$ at physical error rate $p = \SI{1}{\percent}$ on the \SI{22}{\nano\meter} process. The dashed lines indicate the first quartile, median, and third quartile, respectively.}
    \label{fig:violin_latency}
\end{figure}

The distribution of decoding times, as shown in Fig.~\ref{fig:violin_latency}, indicates that the majority of cases complete in a short time.
For example, at $d = 21$, the sandwich decoder completes 90\% of trials within \SI{200}{\nano\second} and 99\% within \SI{500}{\nano\second}.
While $10^4$ trials suffice for the statistics in Table~\ref{tab:execution_cycles}, characterizing the tail requires larger samples.
A qualitative observation by Liyanage et al.~\cite{liyanage2023may} is that syndromes with long decode times tend to have poor decoding accuracy.
Extended evaluation of $10^6$ trials on our sandwich decoder confirmed this at $d = 21$ and $p = \SI{1}{\percent}$, where trials above the 99\% latency percentile exhibit a logical failure rate of 20\%, compared with 0.33\% overall.

Table~\ref{tab:execution_cycles} shows non-monotonic maximum cycle counts at $p = \SI{0.1}{\percent}$.
At $10^6$ trials, however, the 99\% and 99.99\% latency percentiles increase strictly monotonically with $d$ at both error rates, suggesting that the syndrome patterns responsible for the observed maxima are exceedingly rare.

\subsection{Comparison with Existing Hardware Decoders}

Helios~\cite{liyanage2023may} is a distributed Union-Find batch decoder on FPGAs. At $d = 21$ and $p = \SI{0.1}{\percent}$, it achieves an average latency of \SI{11.5}{\nano\second} on a VCU129
FPGA, with computational resources scaling as more than $\order{d^3}$.
The proposed decoder achieves \SI{10.1}{\nano\second} latency on the \SI{22}{\nano\meter} process, and \SI{8.3}{\nano\second} on the \SI{7}{\nano\meter} process at the same code distance and error rate.
Although Helios achieves decreasing average latency with increasing code distance $d$, our proposed decoder has lower latency for all the evaluated code distances from $d = 5$ to $d = 21$.
Additionally, while it is difficult to directly compare the computational resources of FPGA-based and ASIC-based implementations, the circuit area of the proposed decoder scales as $\order{d^3}$, while Helios is expected to scale as $\order{d^3}$ or more, indicating that the proposed decoder is more efficient in terms of computational resources.

Riverlane CC~\cite{barber2025jan} is an improved Union-Find batch decoder for the circuit-level noise model, implemented on an ASIC.
At $d = 23$ and $p = \SI{0.1}{\percent}$, it achieves an average latency of \SI{240}{\nano\second} on a \SI{12}{\nano\meter} FinFET with a circuit area of \SI{0.064}{\milli\meter\squared}.
In contrast, the proposed decoder achieves \SI{9.6}{\nano\second} on the \SI{7}{\nano\meter} process with a circuit area of \SI{1.1875}{\milli\meter\squared} at the same code distance and error rate.
The proposed decoder achieves significantly lower latency than Riverlane CC, by a margin larger than the difference in process technology can explain.
The circuit area of Riverlane CC is smaller than that of the proposed decoder, partly because it employs SRAM macros for graph storage, whereas the proposed decoder uses only standard cell registers.
The implications of adopting SRAM-based storage in the proposed architecture are discussed in Section~\ref{sec:discussion}.

QECOOL~\cite{ueno2021dec} is a batch decoder using the Greedy algorithm implemented on SFQ circuits.
On a \SI{1}{\micro\meter} RSFQ process, at $d = 13$ and $p = \SI{0.1}{\percent}$, it achieves an average latency of \SI{15.3}{\nano\second} with a circuit area of \SI{198.7}{\milli\meter\squared}.
In contrast, the proposed decoder achieves average latencies of \SI{3.5}{\nano\second} with \SI{0.8449}{\milli\meter\squared} on the \SI{22}{\nano\meter} process and \SI{2.9}{\nano\second} with \SI{0.1935}{\milli\meter\squared} on the \SI{7}{\nano\meter} process at the same code distance and error rate.
The proposed decoder outperforms QECOOL in both latency and circuit area.

Online UF~\cite{kasamura2025} is a Union-Find online decoder implemented on an ASIC, which used the same \SI{22}{\nano\meter} and \SI{7}{\nano\meter} processes used in this work for evaluation.
At $d = 23$ and $p = \SI{0.1}{\percent}$, it achieves an average latency of \SI{136.40}{\nano\second} with a circuit area of \SI{3.1522}{\milli\meter\squared} on the \SI{22}{\nano\meter} process, and \SI{102.38}{\nano\second} with \SI{0.7087}{\milli\meter\squared} on the \SI{7}{\nano\meter} process.
The proposed sandwich decoder achieves average latencies of \SI{11.6}{\nano\second} (\SI{22}{\nano\meter}) and \SI{9.6}{\nano\second} (\SI{7}{\nano\meter}), with circuit areas of \SI{5.1166}{\milli\meter\squared} and \SI{1.1875}{\milli\meter\squared}, respectively.
With the same process technology, the proposed decoder achieves significantly lower latency than Online UF, while the trade-off is an increase in circuit area.

We note that the per-round latency of the sandwich decoder is normalized by $d$ throughout this evaluation for direct comparison with the batch decoder. Under continuous operation, the steady-state per-window time budget is closer to $d - 1$ rounds, so the reported sandwich-decoder numbers are marginally optimistic.

\subsection{Discussion}\label{sec:discussion}

The main cost of the proposed decoder is its circuit area, which scales as $\order{d^3}$ with a 1.3--1.5$\times$ overhead over the batch decoder.
The dominant contributor is the graph registers that store per-edge and per-node state (Fig.~\ref{fig:scaling}).
In the inner decoder used in this work~\cite{kasamura2025}, these registers are accessed in parallel by the Grow Phase, which updates all edge weights in a single cycle, and by the Merge Scanner, which performs combinational zero-detection across entire rows.
Replacing them with SRAM macros would require serializing these accesses, fundamentally changing the microarchitecture and trading latency for area.
Exploring this tradeoff is left as future work.
More broadly, the sandwich architecture is complementary to prior decoders rather than replacing them. Existing designs, including SRAM-based implementations such as the CC decoder~\cite{barber2025jan}, can serve as inner decoders and be integrated into the sandwich structure, trading additional area for reduced decoding latency.
Although the datapath is specialized to the rotated surface code and the $\sim$\SI{1}{\micro\second} measurement cycle of superconducting qubits, the temporal-windowing architecture is largely independent of both.
The code structure enters the design only through the decoding graph held in the graph registers. Porting to another code amounts to regenerating this graph and re-tuning the window parameters $(s, b)$, which trade off per-window latency, circuit area, and decoding accuracy, while the microarchitecture remains unchanged. Window-based decoding has indeed been applied to quantum LDPC codes~\cite{maurer2025realtimedecodinggrosscode}.
Superconducting systems impose the most stringent decoding deadline, so platforms with slower syndrome extraction, such as trapped ions and neutral atoms, only relax the real-time constraint.
Evaluating temporal windowing for photonic architectures, e.g., measurement-based computation on the RHG lattice~\cite{raussendorf2007jun}, is left as future work.

\section{Conclusion}\label{chap:futurework}

In this work, we proposed a microarchitecture capable of executing sandwich decoding for improving the speed of QEC decoders toward FTQC, and demonstrated its effectiveness through ASIC implementation.

Simulation and ASIC implementation results show that the proposed method reduces average latency by \SI{35}{\percent} and worst-case latency by \SI{12}{\percent} at code distance $d = 21$ and physical error rate $p = \SI{1.0}{\percent}$ under phenomenological noise, compared with the batch decoder.
The threshold, while slightly lower than that of the batch decoder, was shown to be comparable to existing hardware decoders.
Circuit area increased by a constant factor compared with the batch decoder, demonstrating that sandwich decoding trades computational resources for reduced latency.

These results demonstrate that the proposed method offers a viable new option for high-speed QEC decoders toward FTQC.
Future work includes exploring SRAM-based serialized datapaths to reduce circuit area, investigating alternative window parameters $(s, b)$, sharing decoder computational resources among multiple logical qubits, exploring sandwich decoding with alternative algorithms beyond Union-Find, and implementing spatial sandwich decoding compatible with lattice surgery.

\section*{Acknowledgment}
This work was supported by JST Moonshot R\&D under Grant Number JPMJMS256F and by JST SPRING, Grant Number JPMJSP2108.

\bibliographystyle{IEEEtran}
\bibliography{ref}

\end{document}